\documentclass[a4paper,fleqn,usenatbib]{mnras}

\usepackage[T1]{fontenc}
\usepackage{ae,aecompl}

\usepackage{amsmath}
\usepackage{amsfonts}
\usepackage{amssymb}
\usepackage{graphicx}
\usepackage{multirow,bigdelim}
\usepackage{adjustbox}
\usepackage{placeins}

\def\CN2{\mbox{$C_N^2 \ $}}

\def\CT2{\mbox{$C_T^2 \ $}}

\def\sigmal2{\mbox{$\sigma ^{2}_{I} \ $}}

\title[Forecasting surface atmospheric parameters at LBT]{Forecasting surface layer atmospheric parameters at the LBT site}

\author[A. Turchi et al.]{
Alessio Turchi,$^{1}$\thanks{E-mail: aturchi@arcetri.astro.it}
Elena Masciadri,$^{1}$\thanks{E-mail: masciadri@arcetri.astro.it}
Luca Fini$^{1}$
\\
$^{1}$INAF - Osservatorio Astrofisico di Arcetri, L.go E. Fermi 5, 50125  Florence, Italy\\
}

\date{Accepted 2016 November 03. Received 2016 October 14; in original form 2016 September 02}

\pubyear{2016}

\begin{document}
\label{firstpage}
\pagerange{\pageref{firstpage}--\pageref{lastpage}}
\maketitle

\begin{abstract}
In this paper we quantify the performances of an automated weather forecast system implemented on the Large Binocular Telescope (LBT) site at Mt. Graham (Arizona) in forecasting the main atmospheric parameters close to the ground. The system employs a mesoscale non-hydrostatic numerical model (Meso-Nh). To validate the model we compare the forecasts of wind speed, wind direction, temperature and relative humidity close to the ground with the respective values measured by instrumentation installed on the telescope dome. The study is performed over a large sample of nights uniformly distributed over two years. The quantitative analysis is done using classical statistical operators (bias, RMSE and $\sigma$) and contingency tables, which allows to extract complementary key information, such as the percentage of correct detection (PC) and the probability to obtain a correct detection within a defined interval of values (POD). Results of our study indicate that the model performances in forecasting the atmospheric parameters we have just cited are very good, in some cases excellent: RMSE for temperature is below 1$^{\circ}$ C, for relative humidity is 14\%, for the wind speed is  around 2.5ms$^{-1}$. The relative error of the RMSE for wind direction varies from 9\% to 17\% depending on the wind speed conditions. This work is performed in the context of ALTA (Advanced LBT Turbulence and Atmosphere) Center project, which final goal is to provide forecasts of all the atmospheric parameters and the optical turbulence to support LBT observations, adaptive optics facilities and interferometric facilities.
\end{abstract}




\section{Introduction}
The study presented in this paper was developed in the context of the Advanced LBT Turbulence and Atmosphere (ALTA Center\footnote{\url{http://alta.arcetri.astro.it}}) project, which aims to implement an automated forecast system for the Large Binocular Telescope (LBT) using non-hydrostatic mesoscale atmospheric models. The purpose of ALTA is to perform forecasts of classical atmospheric parameters (wind speed and direction, temperature, relative humidity)
which are relevant for ground-based astronomy and astroclimatic parameters ($C_N^2$ profiles, seeing $\epsilon$, isoplanatic angle $\theta_0$, wavefront coherence time $\tau_0$) to support ground-based observations of LBT. The forecasts of all these parameters are crucial for the telescope operations and are relevant for adaptive optics applications (AO). \\
In the context of this extended project, the goal of this paper is to validate the forecasts of atmospheric parameters (temperature, relative humidity, wind speed and direction) close to the ground above Mount Graham (Arizona), site of the LBT, by comparing outputs of the model (i.e. the forecast system) with measurements of the same parameters (stored in the telescope telemetry) taken by instruments placed on the telescope. More precisely this paper quantifies the confidence level of the model predictions of the parameters just cited.\\
The knowledge in advance of the value of these parameters close to the ground is crucial to maximize the efficiency of the telescope operations and the scheduling of the planned observations. We refer the reader to \cite{masciadri2013} - Section 2 in which it is extensively explained how the atmospheric parameters close to the ground play a fundamental role in optimizing the ground-based observations of telescope facilities, particularly if they are supported by AO systems. 
Here we summarize the main arguments. The dome seeing, one of the main contributions to total seeing, is proportional to the temperature gradients between primary mirror, dome temperature and external temperature, thus knowing in advance the temperature close to the ground is fundamental to minimize the thermal gradients and, as a consequence, the dome seeing. The wind speed close to the ground is the main cause of vibrations of the telescope mirrors. The noise produced by the wind bursts is one of the main causes of error introduced in the AO systems. This effect is mostly visible when the wind hits frontally the mirrors, while it is minimal when hitting laterally. Precise knowledge in advance of the wind direction helps in selecting the suitable part of the sky used for observations, in order to minimize the impact of strong winds on AO systems. Also wind direction is known to be well correlated to seeing conditions. The relative humidity forecast is very useful to be able to close the dome when the RH reaches values larger than the fixed threshold. Of course all these elements, joint with the forecast of the optical turbulence (not analyzed in this paper), will contribute to optimize the scheduling of telescope observations and the telescope scientific outputs.\\

The ALTA Center is integral part of the new strategy conceived by the LBT Observatory (LBTO) to optimize the science operations of the LBT telescope in the near future \citep{veillet2016}. The ALTA project commissioning is articulated in different phases. We intend here with the term `commissioning' the validation of model forecasts or, equivalently, the estimate of the model performances in predicting specific parameters. Results contained in this paper are basically a certification of how good or bad are the model performances with respect to the atmospheric parameters close to the ground. Users and staff responsible for the LBT scheduling can now take advantage of the forecasts of the atmospheric parameters knowing the model performances.

The numerical models used in the ALTA project to forecast the atmospheric paramaters are the Meso-Nh \citep{lafore98} model developed by the Centre National des Recherches M\'et\'eorologiques (CNRM) and Laboratoire d'A\'erologie (LA) and the Astro-Meso-Nh  module \citep{masciadri99a} which is used to provide forecasts of OT parameters. We limit in the context of this paper to the first one.

Among the first studies in using of non-hydrostatic mesoscale models in forecasting atmospheric parameters close to the surface for astronomical applications we highlight (\citealt{masciadri2001, masciadri2003}). At that epoch it was particularly relevant the employment of sub-kilometric horizontal resolutions in the simulations (in the context of the astronomical applications). More recently further attempts followed with the same mesoscale model Meso-Nh \citep{lascaux2009} above the Antarctic plateau of Dome C and with the WRF model \citep{giordano2013}\footnote{This study used only a kilometric horizontal resolution.} above Roque de los Muchchos. More recently, forecasts of atmospheric parameters was part of a large validation campaign conducted within the MOSE project, commissioned by the European Souther Observatory (ESO) and applied to Cerro Paranal and Armazones in Chile using the Meso-Nh model \citet{masciadri2013}. In that context a detailed study has been carried out on the performances of the Meso-Nh model in forecasting the most important atmospheric parameters close to the ground (\citealt{lascaux2013,lascaux2015}). For the first time an exhaustive statistical analysis including statistical operators such as the percent of correct detection and the probability of detection in specific ranges of values has been presented putting in evidence excellent model performances (at our knowledge the best ever achieved in this field) for basically all the parameters.

In this work we aim to perform a model validation study of the surface atmospheric parameters calculated on a sample of 144 nights uniformly distributed between 2014 and 2015. We intend to use the same strategy used for \citet{lascaux2015} but on a different site. This is the first study on the performances of a non-hydrostatic mesoscale model applied to an astronomical site in which the model is already running in an operational system. Besides, we present a characterization of the surface layer atmospheric parameters above the site of LBT. At our knowledge there are no published results in the literature and this is a fundamental first step for our analysis.\\

In Section \ref{overview} we give an overview of the LBT site, measurements characteristics, sample selection criteria and model setup. In Section \ref{char_atm} we present a synthesis of the climatological analysis of the surface atmospheric parameters: temperature, relative humidity, wind speed and direction. In Section \ref{numana} we describe the strategy of the statistical analysis performed in this study. In Section \ref{validation} we show the results of the model validation study, in terms of statistical operators and contingency tables. In Section \ref{concl} we draw the conclusions and perspectives. \\

\section{Numerical analysis overview}
\label{overview}

\subsection{Observations}
\label{measuresatlbt}

\begin{figure*}
\begin{center}
\begin{tabular}{cc}
\includegraphics[height=5cm]{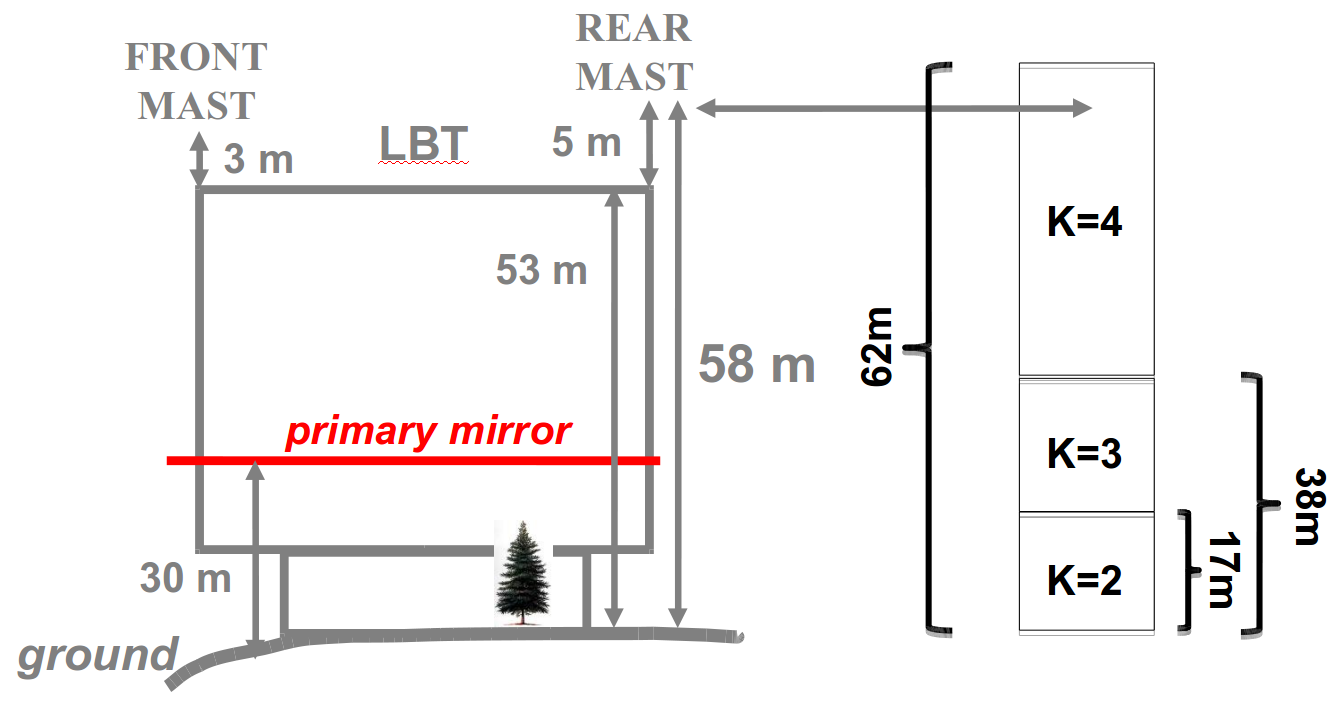} & \includegraphics[height=5cm]{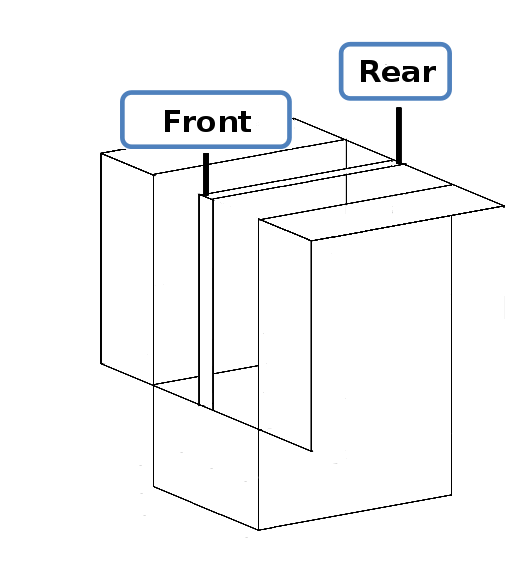}
\end{tabular}
\end{center}
\caption{Large Binocular Telescope dome scheme with reported weather mast locations. We report also the model levels K=2,3,4 with the corresponding heights.}
\label{fig:example_figure}
\end{figure*}

Measurements used as a reference are obtained from the weather stations positioned on two masts placed above the telescope dome and afterwards stored in the telemetry of LBT. Fig. \ref{fig:example_figure} shows telescope and the associated weather station instruments. The LBT dome has a flat roof that is 53~m high above the ground. The mast labeled ``FRONT'' is positioned on the front side of the dome, facing the telescope line of sight direction. This mast has only one anemometer measuring wind speed and wind direction, placed at a height of 3~m above the roof (i.e. 56~m above the ground). The second mast labeled ``REAR'' is placed on the rear side of the dome and is equipped with an anemometer, at 5~m height above the roof (i.e. 58~m above the ground), and a set of sensors measuring temperature, relative humidity and pressure, placed at 2.5~m above the roof (i.e. 55.5~m above the ground). The weather station send data to the telemetry streams approximately each second, which are stored in the LBT archive. We checked that REAR and FRONT anemometer provide the same measure for the wind direction, thus for this specific parameter we used measurements coming from the REAR anemometer for the analysis presented in this paper. In the case of the wind speed we detected discrepancies between the two sensors, depending on the direction of the incoming wind. We could conclude that this was due to the position of the anemometers relative to the LBT roof and we found a method to disentangle the biased measurements and to discard them. We have been forced therefore to use both FRONT and REAR sensors to reconstruct a reliable measure of this parameter. In Section \ref{winddata} we discuss the actual procedure for the wind speed measurements selection.\\

\subsection{Sample selection}
\label{samplesel}
To validate the model we selected a rich sample of 144 nights uniformly distributed between 2014 and 2015. This sample is rich enough to be statistically significative and permits us to perform a reasonable number of simulations with an homogenous model configuration. The selection criterion was therefore the following: starting from 2014/01/01, we selected a date every roughly 5 days with the full night data available. In cases were the telemetry data were missing, we selected the closest night with the full night data available. This criterion permitted us to have around six nights for each month. The selected dates are reported in Table \ref{tab:dates}. We observed that the telemetry archive presented some lack of measurements here and there mainly due to temporary failures of the sensors. Telemetry streams have flags which signal if the corresponding instrument is working and online, however when the night sample was already selected we noticed that there were cases in which the instrument was clearly malfunctioning while the corresponding flag was signaling it was working properly. Specifically we had to remove one night from the relative humidity measurements sample, 5 nights from the wind speed sample and 8 nights from the wind direction sample. We refer to Table \ref{tab:samplesize} for the number of nights used to validate each atmospheric parameter considered in this analysis.\\
Besides we noticed that the relative humidity values extracted from telemetry were distributed between a minimum of 4\% to a maximum of 104\% (exact numbers). Values larger than 100\% are obviously not realistic. We do not know which is the cause of this problem. We therefore considered two cases in our analysis. In case A we assume that the sensor has a bias of + 4\% and we therefore subtract to all measurement 4\%. In case B we simply disregard all measurement between 100\% and 104\%. In both cases measurements are included in the interval [0\%, 100\%]\footnote{For completeness beside cases A and B we might assume a case C in which the problem causing a RH larger than 100\% might affect also the other measurements in an unknown way. However this should be equivalent to assume that all measurements are not reliable at all. This does not seems to be the case as we will see later.}.
\begin{table*}
\begin{center}
\caption{Selected nights (in UT time, yyyy/mm/dd) for preliminary model validation.} 
\label{tab:dates}
\begin{tabular}{|c|c|c|c|c|c|c|c|}
\hline
 2014/01/01 & 2014/01/06 & 2014/01/11 & 2014/01/16 & 2014/01/21 & 2014/01/26 & 2014/02/01 & 2014/02/05 \\
 2014/02/11 & 2014/02/16 & 2014/02/21 & 2014/02/26 & 2014/03/01 & 2014/03/06 & 2014/03/11 & 2014/03/16 \\
 2014/03/21 & 2014/03/26 & 2014/04/01 & 2014/04/06 & 2014/04/11 & 2014/04/16 & 2014/04/21 & 2014/04/26 \\
 2014/05/01 & 2014/05/06 & 2014/05/11 & 2014/05/15 & 2014/05/21 & 2014/05/26 & 2014/06/01 & 2014/06/06 \\
 2014/06/11 & 2014/06/16 & 2014/06/21 & 2014/06/26 & 2014/07/01 & 2014/07/06 & 2014/07/10 & 2014/07/16 \\
 2014/07/22 & 2014/07/26 & 2014/08/01 & 2014/08/06 & 2014/08/11 & 2014/08/16 & 2014/08/19 & 2014/08/28 \\
 2014/09/01 & 2014/09/06 & 2014/09/11 & 2014/09/16 & 2014/09/21 & 2014/09/26 & 2014/10/01 & 2014/10/06 \\
 2014/10/11 & 2014/10/16 & 2014/10/21 & 2014/10/26 & 2014/11/01 & 2014/11/06 & 2014/11/11 & 2014/11/16 \\
 2014/11/21 & 2014/11/26 & 2014/12/01 & 2014/12/06 & 2014/12/11 & 2014/12/16 & 2014/12/21 & 2014/12/26 \\
 2015/01/01 & 2015/01/06 & 2015/01/10 & 2015/01/16 & 2015/01/21 & 2015/01/26 & 2015/02/04 & 2015/02/06 \\
 2015/02/11 & 2015/02/16 & 2015/02/21 & 2015/02/26 & 2015/03/06 & 2015/03/07 & 2015/03/11 & 2015/03/16 \\
 2015/03/21 & 2015/03/26 & 2015/04/01 & 2015/04/06 & 2015/04/11 & 2015/04/16 & 2015/04/21 & 2015/04/28 \\
 2015/05/06 & 2015/05/07 & 2015/05/11 & 2015/05/18 & 2015/05/21 & 2015/05/26 & 2015/06/01 & 2015/06/06 \\
 2015/06/11 & 2015/06/15 & 2015/06/21 & 2015/06/26 & 2015/07/01 & 2015/07/06 & 2015/07/11 & 2015/07/16 \\
 2015/07/21 & 2015/07/26 & 2015/08/14 & 2015/08/16 & 2015/08/17 & 2015/08/18 & 2015/08/21 & 2015/08/24 \\
 2015/09/02 & 2015/09/06 & 2015/09/12 & 2015/09/16 & 2015/09/21 & 2015/09/26 & 2015/10/10 & 2015/10/16 \\
 2015/10/24 & 2015/10/25 & 2015/10/26 & 2015/10/28 & 2015/11/03 & 2015/11/08 & 2015/11/10 & 2015/11/14 \\
 2015/11/24 & 2015/11/25 & 2015/12/05 & 2015/12/11 & 2015/12/14 & 2015/12/19 & 2015/12/21 & 2015/12/30 \\
\hline
\end{tabular}
\end{center}
\end{table*}

\begin{table*}
\begin{center}
\caption{Total number of nights used for the validation of each atmospheric parameter, together with the nights excluded from the total 144 nights sample from Table \ref{tab:dates}.}
\label{tab:samplesize}
\begin{tabular}{|c|c|c|} 
\hline
  \textbf{Parameter} & \textbf{Number of nights} & \textbf{excluded nights}\\
\hline
  Temperature & 144 & \\
\hline
  Relative humidity & 143 & 2015/09/21 \\
\hline
  Wind speed & 139 & 2014/02/01 2014/03/01 2014/11/16 2014/12/26 2015/01/01  \\
\hline
  Wind direction & 136 & 2014/02/01 2014/03/01 2014/07/16 2014/07/22 \\
   &  &2014/11/16 2014/12/26 2015/01/01 2015/03/26 \\
\hline
\end{tabular}
\end{center}
\end{table*}

\subsection{Model configuration}
\label{modelconf}
The numerical model used to produce the forecasts of the aforementioned parameters is Meso-Nh\footnote{\url{http://mesonh.aero.obs-mip.fr/mesonh/}}, which is an atmospheric nonhydrostatic mesoscale model that simulates the time evolution of weather parameters in a three-dimensional volume over a finite geographical area. The coordinates system is based on Mercator projection, which is most suitable at low latitudes as in the LBT case, while the vertical levels use the Gal-Chen and Sommerville coordinate systems \citep{chen}.
The model filter acoustic waves thanks to an anelastic formulation of hydrodynamic equations. Simulations are made using a one-dimensional mixing length proposed by Bougeault and Lacarrere \citep{Bougeault89} with a one-dimensional 1.5 closure scheme \citep{Cuxart00}. The exchange between surface and atmosphere is computed with the Interaction Soil Biosphere Atmosphere - ISBA scheme \citep{Noilhan89}.

The model is initialized with forecast data provided by the European Center for Medium Weather Forecasts (ECMWF), calculated with their hydrostatic General Circulation Model (HRES) extend on the whole globe, with an horizontal resolution of 16~km\footnote{Since March 2016 resolution increased to roughly 9~km.}. The Meso-Nh model initialization starts each night at 00:00 UT (i.e. 17:00 MST) and the Meso-Nh model is forced every six hours with new date from the ECMWF. We simulate a total of 15 hours up to 15:00 UT (i.e. 08:00 MST) in order to completely cover the night time period of Mt. Graham. This is exactly the procedure used in the operational configuration.

\begin{figure}
\begin{adjustbox}{max width=\columnwidth}
\begin{tabular}{cc}
\includegraphics[height=5cm]{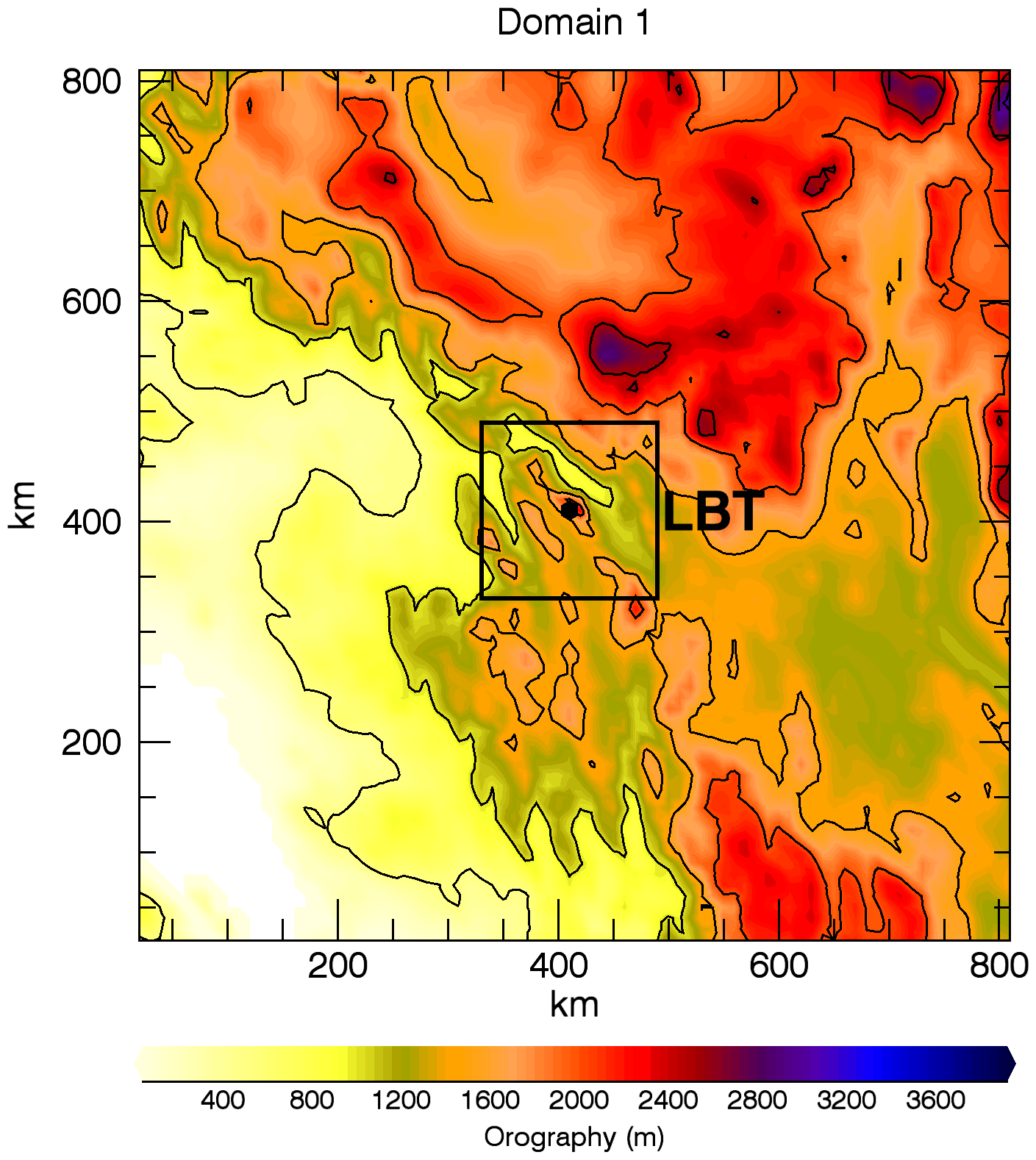} & \includegraphics[height=5cm]{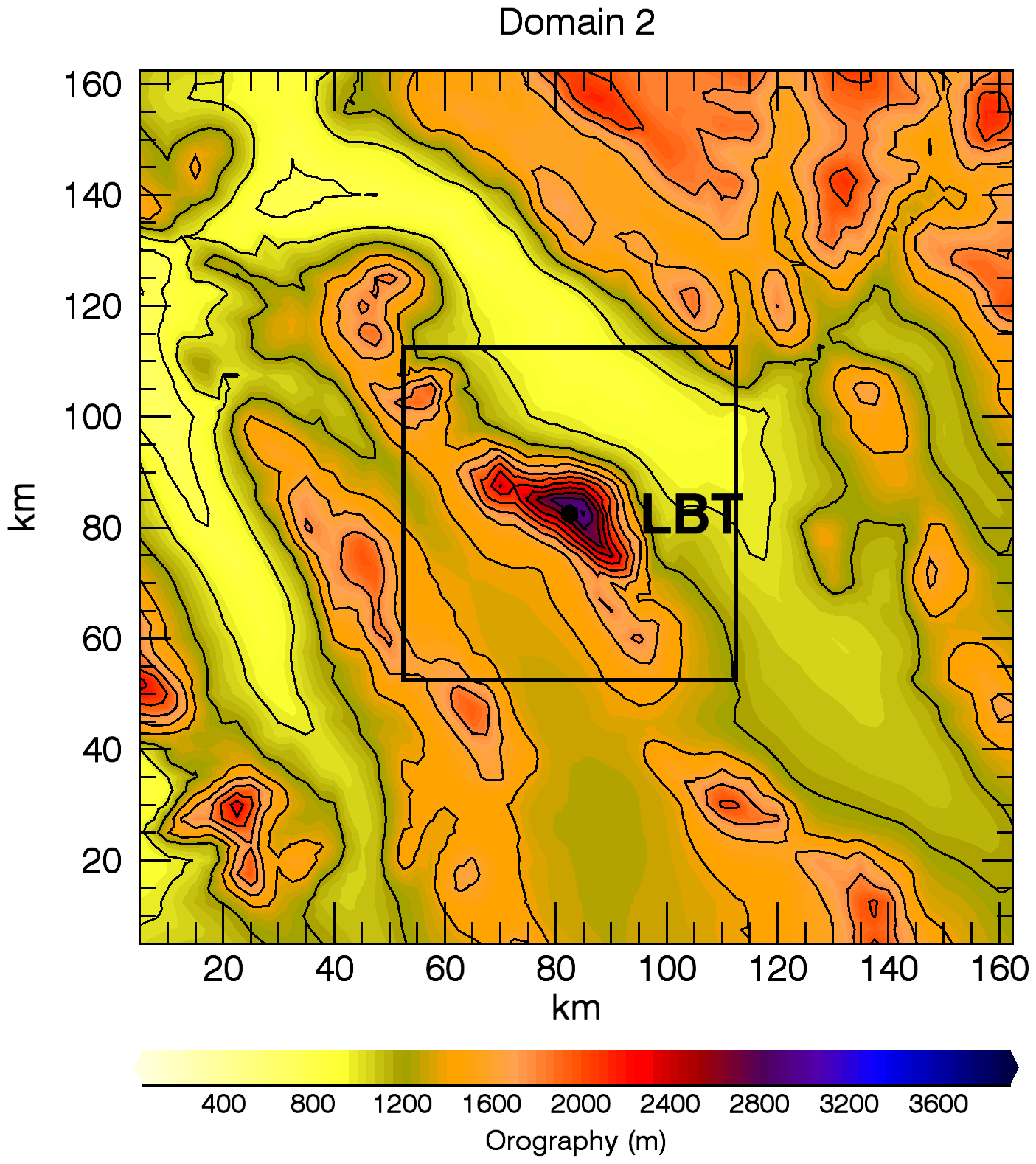}\\
\includegraphics[height=5cm]{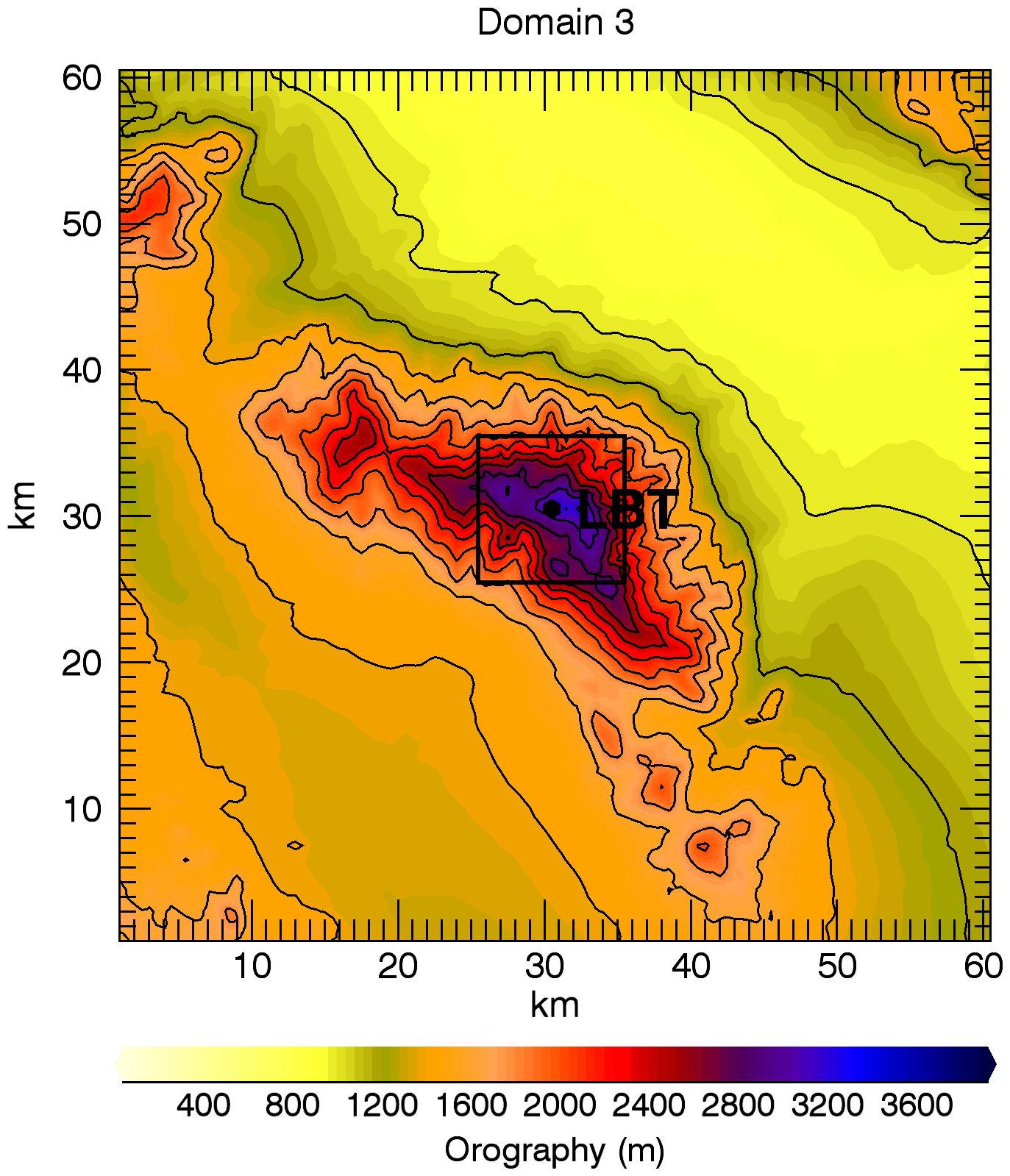} & \includegraphics[height=5cm]{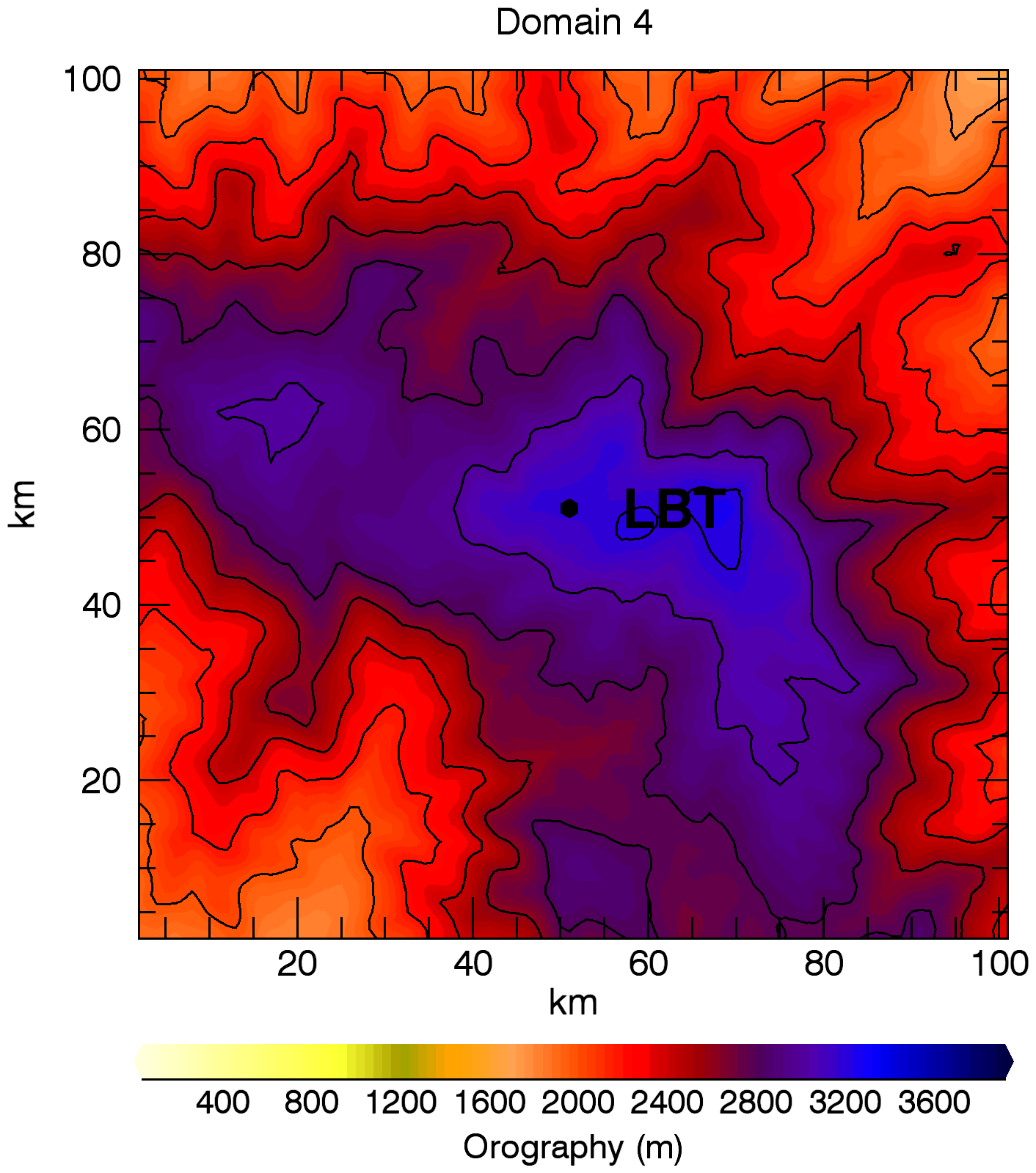}\\
\end{tabular}
\end{adjustbox}
\caption{Horizontal maps of model nested domains, from the outer to the innermost domain, as reported in Table \ref{tab:resol}. The black dot in the center of each domain corresponds to the position of LBT, while the color scale represents the orography elevation.}
\label{nesting}
\end{figure}

The Mt. Graham site of LBT is located at coordinates $[32.70131, -109.88906]$, at an height of 3221~m above sea level (a.s.l.). We used a grid-nesting technique \citep{Stein00}, that consists of using different imbricated domains, described in Table \ref{tab:resol}, with digital elevation model (DEM, i.e. orography) extended on smaller and smaller surfaces having a progressively higher horizontal resolution. This procedure consents to achieve a high horizontal resolution, using the same vertical grid resolution, on a sufficiently small region around the summit of interest to provide better model predictions at the specific site. Each domain is centered on the LBT coordinates. A graphic representation of the model domains is given in Fig. \ref{nesting}. \cite{lascaux2013} proved that an horizontal resolution of 100~m is necessary at Cerro Paranal and Cerro Armazones to reconstruct wind speed close to the ground when the wind speed is strong. In our study we compare the results obtained with 500~m resolution with the ones obtained with 100~m resolution for the wind speed to verify if similar conditions are found also above Mt. Graham. In the case of the other atmospheric variables (temperature, relative humidity, wind direction), we used a resolution of 500~m that is sufficient to reconstruct reliable values of these parameters.

\begin{table}
\begin{center}
\caption{Horizontal resolution of each Meso-Nh imbricated domain.}  
\begin{tabular}{cccc} 
\hline
  Domain & $\Delta$X (km) & Grid points & Domain size (km) \\
\hline
  Domain 1 & 10 & 80x80 & 800x800 \\
  Domain 2 & 2.5 & 64x64 & 160x160 \\
  Domain 3 & 0.5 & 120x120 & 60x60 \\
  Domain 4 & 0.1 & 100x100 & 10x10 \\
\hline
\end{tabular}
\label{tab:resol}
\end{center}
\end{table}
For what concerns the orography on domains 1 and 2, we used the GTOPO\footnote{\url{https://lta.cr.usgs.gov/GTOPO30}} DEM, with an intrinsic resolution of 1~km. In domains 3 and 4 we utilized the SRTM90\footnote{\url{http://www.cgiar-csi.org/data/srtm-90m-digital-elevation-database-v4-1}} DEM \citep{jarvis2008}, with an intrinsic resolution of approximately 90~m (3 arcsec). The resolution is obviously defined at the end by the number of grid-points of the atmospheric model therefore 500 and 100 meters in our case. \\
For all the couples of imbricated domains we uses a two-way interacting grid-nesting, in which the interface between outer and inner domain has a bidirectional interaction. This allows the atmospheric flow between each domain to be in constant thermodynamic equilibrium with the outer one, consenting for the propagation of gravity waves through the whole area mapped by the simulation independently on the specific domain.

The simulation has 54 physical vertical levels on each domain, with the first grid point set to 20~m above ground level (a.g.l.), and a logarithmic stretching of 20\% up to 3.5~km a.g.l following Eq.\ref{eq:log}:
\begin{equation}
\frac{\Delta z(k+1)}{\Delta z(k)}=1+\frac{20}{100}
\label{eq:log}
\end{equation}
From this point onward the model uses an almost constant vertical grid size of $\sim$ 600~m up to 23.57 km, which is the top level of our domain. 

As reported in section \ref{measuresatlbt}, LBT weather stations are positioned $\sim $ 55-58~m above ground, and this corresponds to the third physical Meso-Nh level (K=4, Fig. \ref{fig:example_figure}) spanning the interval [38-62]~m at LBT site coordinates. The model output for each level from the innermost model is representative of the whole interval spanned by the level itself.\\
The model has been configured in order to give access to the temporal evolution of the surface parameters calculated on the summit (LBT location) with a temporal frequency equal to the time step (order of a few seconds) of the innermost domain.

\section{Climatologic characterization of the surface layer atmospheric parameters}
\label{char_atm}

\begin{figure*}
\centering
\includegraphics[width=0.8\textwidth]{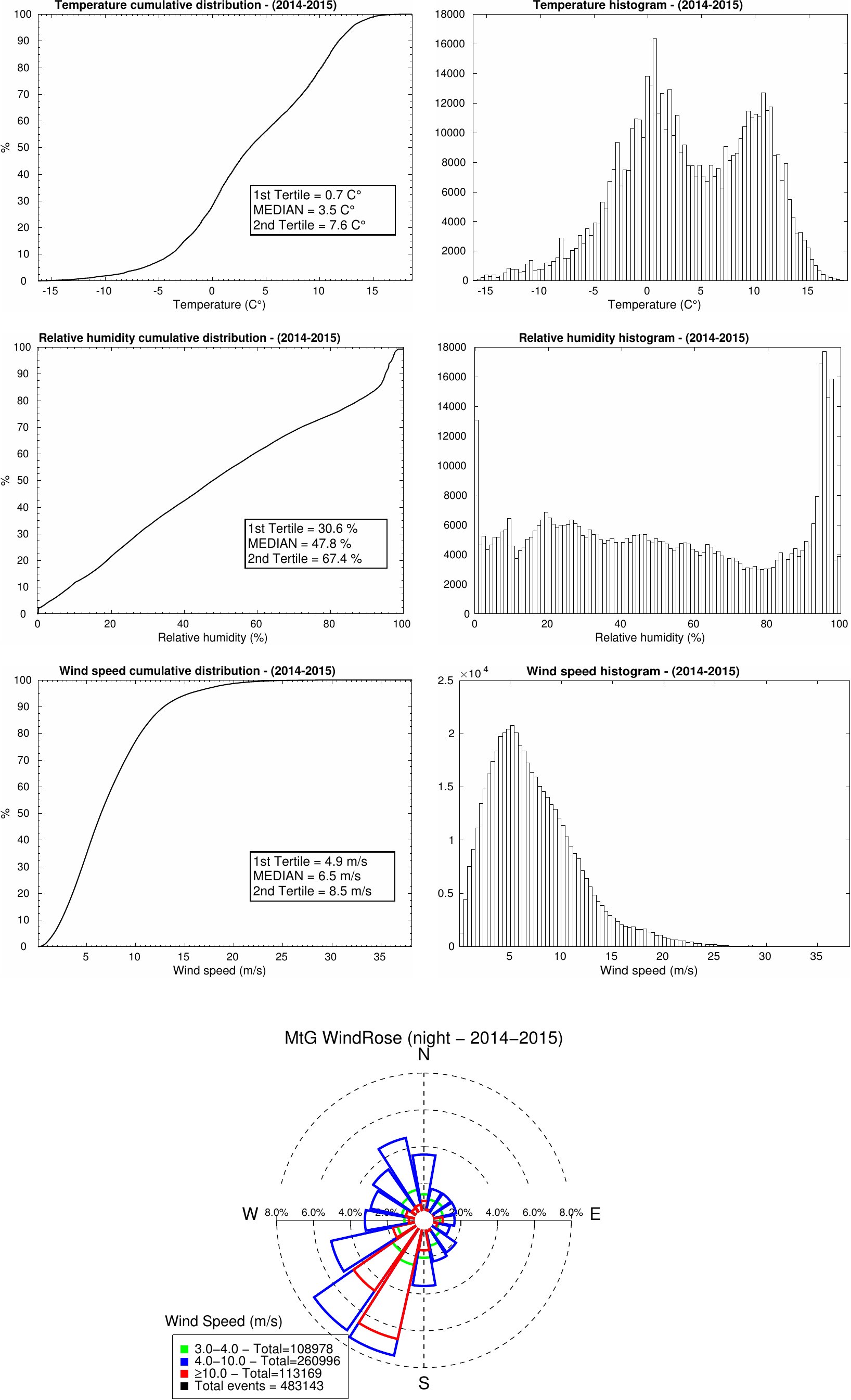}
\caption{Cumulative distributions, histograms and wind rose computed on the whole 2014-2015 years, considering measurements between sunrise and sunset.}
\label{fig:tempcumdist1}
\end{figure*}

\begin{figure*}
\centering
\includegraphics[width=0.8\textwidth]{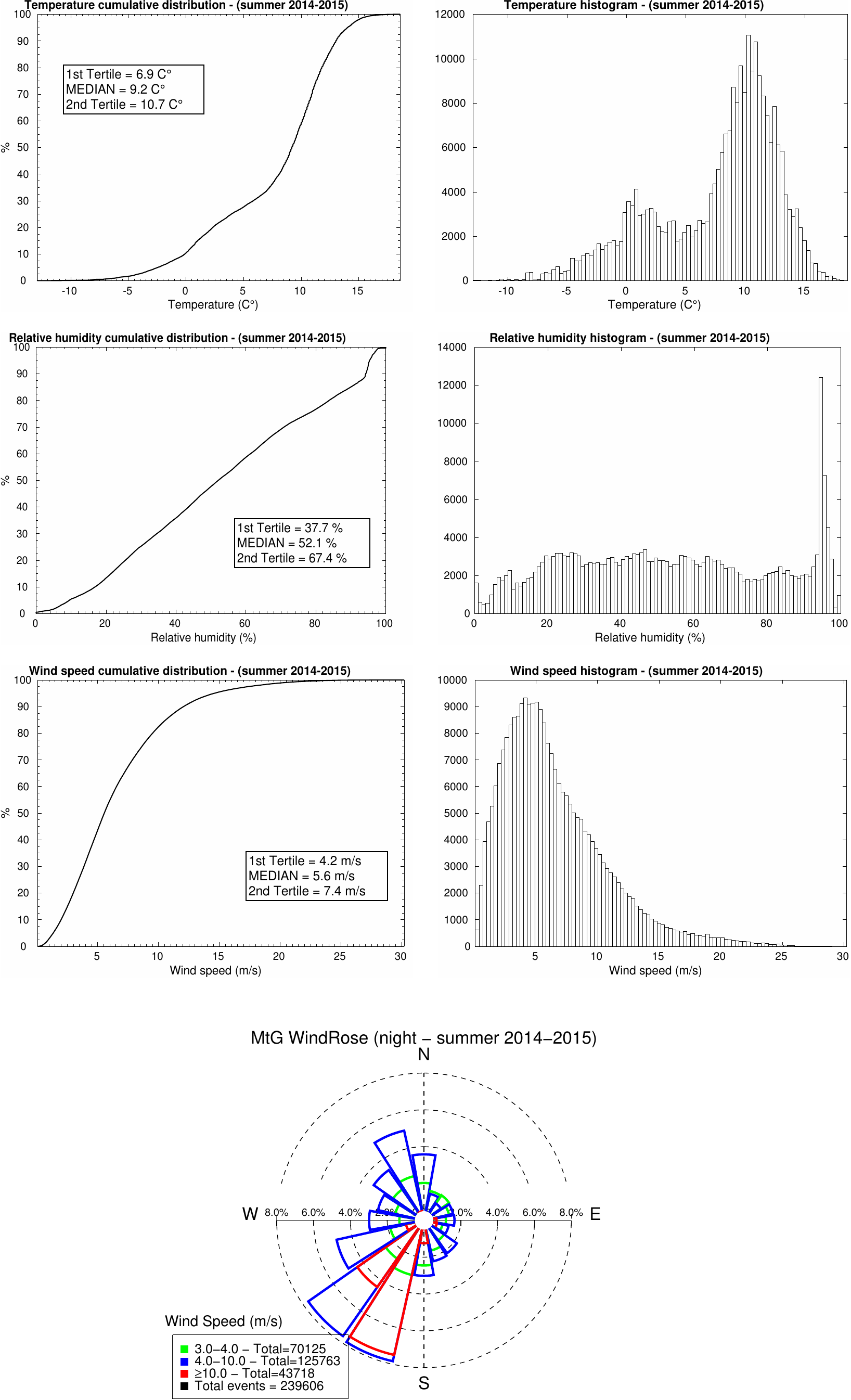}
\caption{Cumulative distributions, histograms and wind rose computed on the summer time (April-September) of 2014-2015 years, considering measurements between sunrise and sunset.}
\label{fig:tempcumdist2}
\end{figure*}

\begin{figure*}
\centering
\includegraphics[width=0.8\textwidth]{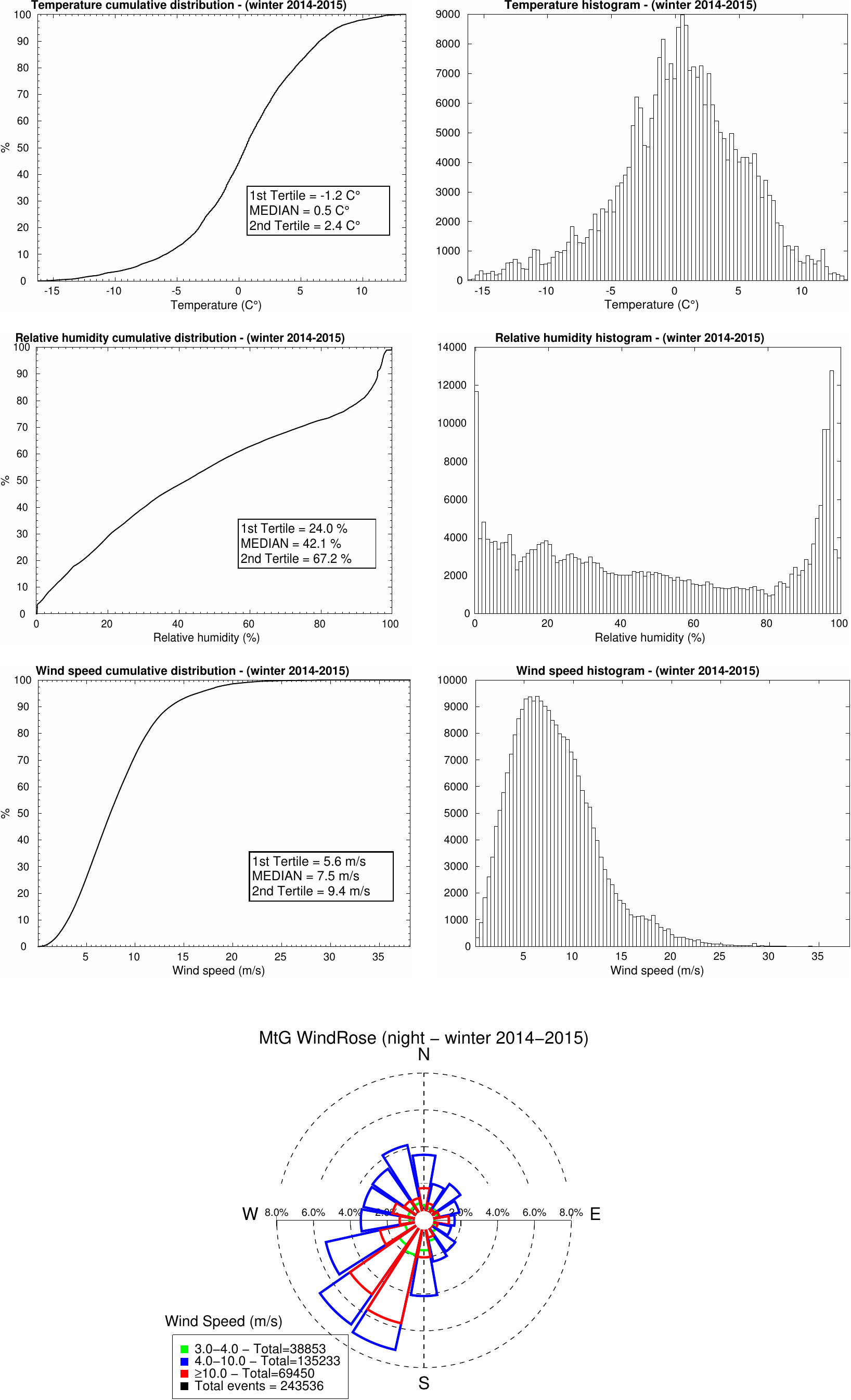}
\caption{Cumulative distributions, histograms and wind rose computed on the winter time (October-March) of 2014-2015 years, considering measurements between sunrise and sunset.}
\label{fig:tempcumdist3}
\end{figure*}

We report in this section the climatology distributions of the atmospheric parameters (temperature, wind speed and direction, relative humidity) in the surface layer at the summit of Mt. Graham. At our knowledge there are no published results on the characterization of these parameters above Mt.Graham. This should be therefore the first one in this respect. We report the cumulative distribution and histogram for each parameter, either for the full 2014-2015 years (730 nights) considered in this analysis (Fig. \ref{fig:tempcumdist1}) and for the summer (April-September, Fig. \ref{fig:tempcumdist2}) and winter (October-March, Fig. \ref{fig:tempcumdist3}) periods of the same date sample. For each date we extract the subsample of data included between sunset and sunrise. After the above procedure the distributions are computed. Tertiles and median values are calculated for each parameter.

In Fig. \ref{fig:tempcumdist1} we observe that temperature distribution has two distinct peaks at around $\sim 10^\circ C$ and $\sim 1^\circ C$. By observing the partial temperature distributions in Fig. \ref{fig:tempcumdist2} and \ref{fig:tempcumdist3} relative to summer and winter, we can easily detect that the two temperature peaks are related to the median temperature in summer and winter respectively. However in Fig. \ref{fig:tempcumdist3} it is evident that there is a secondary small peak at around $\sim 1^\circ C$ meaning that even in summer time there are events where outside temperature is near or below the freezing point. The relative humidity (RH) distribution show that there are numerous events, both in summer and winter, where the values are near $\sim100\%$, mostly related to rainy events. However the driest season is winter, while summer tend to have a more flat distribution of relative humidity values. Median value of RH on the whole year is 47.8\%, in winter is 42.1\% and in summer is 52.1\%. \\
Regarding the wind speed distribution, we observe that the median value on the whole year is 6.5 ms$^{-1}$ while in winter is $\sim7.5$ ms$^{-1}$ and in summer is $\sim5.6$ ms$^{-1}$. The percent of cases where wind speed is larger than $15$ ms$^{-1}$ is 6\% (with no major differences between winter and summer) and those larger than 20 ms$^{-1}$ is 1\%.\\
The wind direction distribution does not show much seasonal variability. From its distribution we observe that strong winds (above 10 ms$^{-1}$, those most critical for astronomers) are mainly coming from south-west direction, however there is always a chance to observe strong wind from all directions.\\


%

\section{Strategy of analysis}
\label{numana}
As already done in recent studies (\citealt{lascaux2013,lascaux2015}) on Cerro Paranal and Cerro Armazones\footnote{Decision taken in agreement with the ESO and LBTO staff.} we decided to perform a moving average over a 1-hour time window, from 30 minutes before to 30 minute after, both on measurements and numerical forecasts. This operation allows for the filtering of fast frequencies and consents us to estimate the model performance over the slower-moving trends, which are of interest for the telescope operation and planning. Data were then resampled over 20-minutes intervals. Such a value has been selected because the time required to switch a beam from an instrument to another one (or to a program to another one) is typically of the order of 20 minutes. It makes therefore no sense to use a higher frequency. We highlight the fact that, even if we calculate a moving average that smooth out the high frequencies it is extremely important to have an high frequency output from the model (as it is our case) instead of treating outputs with a temporal sampling of 1 hour (as is the case in many studies in the literature) since it allows for a more reliable estimate of the atmospheric quantities over the selected time window. Finally, we selected only values contained between sunset an sunrise hours, computed with ephemerids tables for each date.\\
In Table \ref{tab:sizesun} we report the total sample size (number of couples of points ($p_{i,sim}$, $p_{i,obs}$)) considered for each atmospheric parameter, once the procedure described in this section was applied. For the wind direction we considered different samples in which we filtered out values which corresponds to associated wind speeds lower than 3, 5 and 10 ms$^{-1}$ (see discussion in Section \ref{winddir}).\\

\begin{table}
\begin{center}
\caption{Total sample size (couple of points ($p_{i,sim}$, $p_{i,obs}$) for each atmospheric parameter related to the total 144 nights sample from Table \ref{tab:dates}. For the wind direction we considered different samples because we filtered out values associated to a wind speed larger than 3, 5 and 10 ms$^{-1}$.}
\begin{tabular}{|c|c|} 
\hline
  \textbf{Parameter} & \textbf{Sample size} \\
\hline
  Temperature & 5201 \\
\hline
  Relative humidity & 5165 \\
\hline
  Wind speed & 4996 \\
\hline
  Wind direction (WS>3~ms$^{-1}$) & 4423 \\
  Wind direction (WS>5~ms$^{-1}$) & 3626 \\
  Wind direction (WS>10~ms$^{-1}$) & 1311 \\
\hline
\end{tabular}
\label{tab:sizesun}
\end{center}
\end{table}

The validation procedure used in this paper, similarly to what has already been done for ESO telescope sites by \citep{lascaux2015}, followed two different statistical approaches. First we computed the classical statistical operators bias, RMSE and $\sigma$, defined as follows:
\begin{equation}
BIAS = \sum\limits_{i = 1}^N {\frac{{(Y_i  - X_i )^{} }}
{N}} 
\label{eq1}
\end{equation}
\begin{equation}RMSE = \sqrt {\sum\limits_{i = 1}^N {\frac{{(Y_i  - X_i )^2 }}
{N}} } 
\label{eq2}
\end{equation}
where $X_{i}$ are the individual observations and $Y_{i}$ the individual numerical forecasts calculated at the same time index $i$ nd $N$ being the total sample size.\\
From the above quantities we deduce the bias-corrected RMSE ($\sigma$):
\begin{equation}\sigma = \sqrt {RMSE^2 - BIAS^2}
\label{eqr}
\end{equation}
The previously defined indicators provide us a global information on the statistical and systematic errors of the model.\\

In order to refine the statistical analysis and have a more practical estimate of the model score of success, we proceeded to perform also a different analysis based on contingency tables, which are able to provide us complementary informations, with respect to the above statistical operators, on the reliability of the model in a realistic use-case scenario. A contingency table is a method to study the relationship between two or more categorical variables and provide multiple statistical operators: the percent of correct detection (PC), the probability to detect a parameter within a specific range of values (POD$_i$) and the probability of extremely bad detection (EBD). We refer to \cite{lascaux2015} for an exhaustive discussion on contingency tables and definition of all these statistical operators.

For temperature, wind speed and relative humidity we decided to use as categories the tertiles of the cumulative distribution of these parameters (3$\times$3 contingency tables) in which the thresholds are the tertiles of a climatological analysis performed on measurements (see Table \ref{tab:tertemp}). An exception was made for the wind direction, which was divided into quadrants defined as follows: North=$[315^\circ, 45^\circ]$, East=$[45^\circ,135^\circ]$, South=$[135^\circ,225^\circ]$, West=$[225^\circ,315^\circ]$ (4$\times$4 contingency table). As can be seen in Section \ref{validation} the selected sample of nights (144) on which we performed the model validation cover all the conditions revealed by the climatological analysis. In other words there are no biased effects.  \\ 

\begin{table}
\begin{center}
\caption{Climatological distribution for atmospheric parameters at Mount Graham. Left column: fist tertile ($33\%$), central column: median ($50\%$), right column: second tertile ($66\%$). Values are computed over the 2014-2015 distribution of observations from LBT telemetry.}
\begin{tabular}{c|ccc}
\hline
 Mount Graham & $33\%$ & median $(50\%)$   &  $66\%$ \\
\hline
Temperature $(^\circ C)$ & 0.7 & 3.5 & 7.6\\
Relative humidity $(\%)$   & 30.6 & 47.8 & 67.4\\
Wind speed (ms$^{-1}$)    & 4.9 & 6.5 & 8.5\\
\hline
\end{tabular} 
\label{tab:tertemp}
\end{center}
\end{table}

\section{Model validation}
\label{validation}
In this section we report the results obtained with the procedure described in section \ref{numana}. The validation is performed over the full sample of nights distributed over the whole two years 2014-2015, with actual sample sizes for each parameter reported in Tables \ref{tab:dates} and \ref{tab:samplesize}. 
Here we report the RMSE, bias and $\sigma$ obtained on the whole sample as well as the contingency tables with associated PC, PODs and EBDs values. The cumulative distributions of RMSE, bias and $\sigma$ errors obtained considering each single night are reported in appendix \ref{appendixcum}. When not otherwise specified, model outputs are taken from the vertical level K=4 (see Fig. \ref{fig:example_figure}), which is representative of the weather masts positions and in the grid point correspondent to the summit of Mt. Graham in the innermost domain having a horizontal resolution of 500~m for temperature, RH and wind direction and a resolution of 100~m for the wind speed. \\

\subsection{Temperature}
\begin{figure}
\begin{center}
\includegraphics[width=6cm]{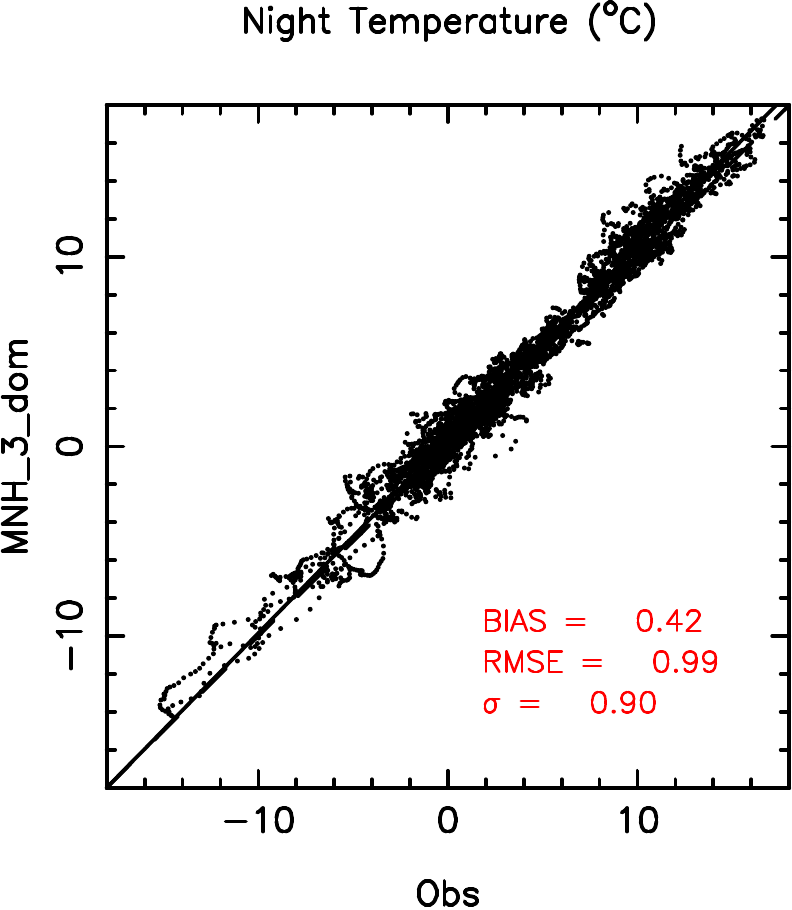}
\end{center}
\caption[example]{\label{fig:treg}Scatter plot for temperature, comparing model outputs (MNH) and measurements (OBS). The full black line is the regression line passing by the origin, while the dashed line represent the reference diagonal line for unbiased results.} 
\end{figure}

In Fig. \ref{fig:treg} is reported the scattering plot between observations and simulations calculated on the sample in Table \ref{tab:samplesize}. We can observe that all the statistical operators are well below the degree Celsius (bias = 0.42$^{\circ}C$, RMSE = 0.99$^{\circ}$ and $\sigma=0.9^{\circ}C$) and this indicates an excellent model performances. 
Table \ref{tab:tempc} is the contingency table obtained for the temperature parameter where the thresholds are the climatological tertiles defined in Table \ref{tab:tertemp}. We observe that the model has an impressive PC=91.7\% and a zero EBD, as well as excellent POD$_{i}$ (between 84\% and 99\%). These results confirm an excellent model performance similar to what we found in a recent study carried out above Cerro Paranal on a sample of 129 nights \citep{lascaux2015}. 

If we consider a sample of bias, RMSE and $\sigma$ calculated for each night and we calculate the cumulative distribution (see Annex \ref{appendixcum}) we observe that the median value of the bias, RMSE and $\sigma$ of the temperature is well below the degree Celsius too. This means that also if we look at the problem from the single nights perspective, the results are equivalently good.

If we consider the two separate sub-samples of summer and winter time we find an equivalent good model performances. Fig. \ref{fig:scattersummer} and Fig. \ref{fig:scatterwinter} (left panel) in Annex \ref{appendix3} show the scattering plots of temperature in summer and winter time, while Table \ref{tab:tempsummerc} and Table \ref{tab:tempwinterc} show the contingency tables calculated using, as thresholds, the tertiles associated to the respective sub-samples. We can observe that the model performances remain good with no major differences observed in the two periods of the years. POD$_{i}$ are well above 80\% with the unique exception of POD$_{2}$ in summer that is 70.1\% that is in any case still very good.

\begin{table}
\begin{center}
\caption{3$\times$3 contingency table for the absolute temperature during the night, at 55.5~m a.g.l. at LBT, for the sample of 144 nights.
We use the Meso-Nh ${\Delta}$X~=~500~m configuration.}
\resizebox{\columnwidth}{!}{
\begin{tabular}{cc|ccc}
 \hline
 \multicolumn{2}{c}{Temperature ($^{\circ}C$)} & \multicolumn{3}{c}{\bf OBSERVATIONS}\\
 \multicolumn{2}{c}{} & ~~T<0.7 &   ~~0.7<T<7.6~~  &  ~~T>7.6 \\
 \hline
 \multirow{7}{*}{\rotatebox{90}{\bf MODEL}} & & &\\
  & T<0.7 &    1381      &    79      &    0     \\
  & & & & \\
  &   0.7<T<7.6 &    267      &    1841      &    20     \\
  & & & & \\
  & T>7.6 &    0      &    67      &    1546     \\
  & & & & \\
 \hline
 \\
 \multicolumn{5}{l}{Sample size = 5201; PC=91.7\%; EBD=0.0\%} \\
  \multicolumn{5}{l}{POD$_1$=83.8\%; POD$_2$=92.7\%; POD$_3$=98.7\%} \\
\end{tabular}
}
\label{tab:tempc}
\end{center}
\end{table}

\subsection{Relative humidity}
\begin{figure}
\begin{center}
\includegraphics[width=6cm]{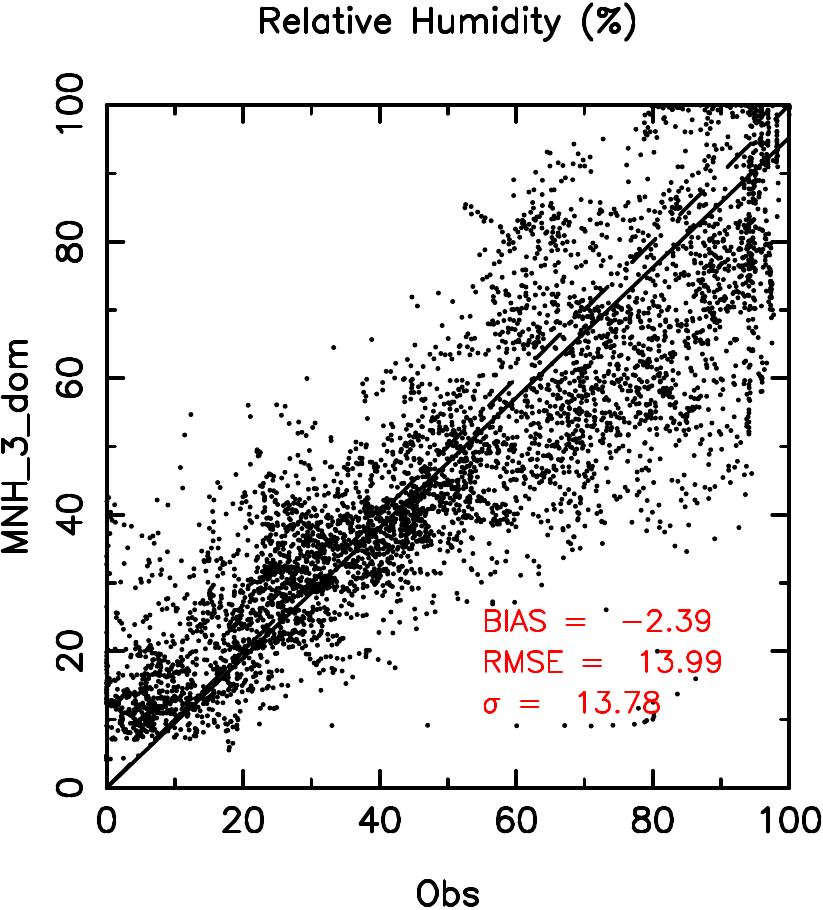}
\end{center}
\caption{Scatter plot for relative humidity, comparing model outputs (MNH) and measurements (OBS). The full black line is the regression line passing by the origin, while the dashed line represent the reference diagonal line for unbiased results.}
\label{fig:rhreg}
\end{figure}

Relative humidity (RH) is the ratio (expressed in percentage) of the mixing ratio $w$ to the saturation mixing ratio $w_s$ with respect to water at the same temperature and pressure: $RH = \frac{w}{w_s} \times 100$. Fig. \ref{fig:rhreg} reports the scattering plot of RH on the sample in Table \ref{tab:samplesize}, as done for the temperature. Values of bias = -2.4\%, RMSE = 14.0\% and $\sigma$ = 13.8\% can be considered absolutely satisfactory. The points are well distributed along the regression line passing by the origin. We note that the dispersion of the cloud points increases with RH values. If we look at the contingency tables (Table \ref{tab:rhc}) we observe a good global PC=71.8\% and EBD=0.3\%. For what concerns the POD$_{i}$ we find good values for POD$_{1}$ = 74.4\% and POD$_{2}$ = 79.6\%. POD$_{3}$ decreases to 61\%. This matches with the increasing dispersion of the scattering plot for large RH (see Fig. \ref{fig:rhreg}). We note that POD$_{3}$ is the most interesting from an astronomical point of view. Astronomers are indeed interested in knowing when the RH is higher than the threshold that permits to keep open the dome for observations. In spite of a more modest performance 61\% is still well above the 33\% value of the random case. 
We highlight that these results have been obtained considering the case A (see Section \ref{samplesel}). We verified that the case B provides negligible differences in results on the scattering plots and contingency table. The only interesting thing is that POD$_{3}$ is 56.2\% instead of 61\%. In conclusions POD$_{3}$ is in the [56.2\%-61\%] range.

We observe that the rich sample of 144 night provides a much better result [56.2\%-61\%] with respect to what we had found in a preliminary study \citep{turchi2016} in which POD$_{3}$ calculated on a small sample of 22 nights was 30\%. 


If we consider a sample of bias, RMSE and $\sigma$ calculated for each night and we calculate the cumulative distribution (see Annex \ref{appendixcum}) we observe that the median value of the bias, RMSE and $\sigma$ are as good as those obtained on the whole sample.

Looking at the scattering plots of the sub-sample of summer and winter (Fig. \ref{fig:scattersummer} and \ref{fig:scatterwinter}, second panel) we do not observe any substantial difference between the periods. Also the contingency tables (Table \ref{tab:rhsummerc} and Table \ref{tab:rhwinterc}) in summer and winter confirm a substantial equivalent behavior of the model with similar performances. POD$_{3}$ in summer is 62.8\% and in winter is 59.4\% (while POD$_{3}$ is 61\% on the total sample).

\begin{table}
\begin{center}
\caption{3$\times$3 contingency table for the relative humidity during the night, at 55.5~m a.g.l. at LBT, for the sample of 143 nights.
We use the Meso-Nh ${\Delta}$X~=~500~m configuration.}
\label{tab:rhc}
\resizebox{\columnwidth}{!}{
\begin{tabular}{cc|ccc}
 \hline
 \multicolumn{2}{c}{Relative humidity (\%)} & \multicolumn{3}{c}{\bf OBSERVATIONS}\\
 \multicolumn{2}{c}{} & RH$<$30.6 & 30.6$<$RH$<$67.4 & RH$>$67.4 \\
 \hline
 \multirow{7}{*}{\rotatebox{90}{\bf MODEL}} & & &\\
  & RH$<$30.6 & 1199 & 183 & 14 \\
  & & & & \\
  &    30.6$<$RH$<$67.4 & 412 & 1475 & 649 \\
  & & & & \\
  & RH$>$67.4 & 0 & 196 & 1037 \\
  & & & & \\
 \hline
 \\
 \multicolumn{5}{l}{Sample size = 5165; PC=71.8\%; EBD=0.3\%} \\
  \multicolumn{5}{l}{POD$_1$=74.4\%; POD$_2$=79.6\%; POD$_3$=61.0\%} \\
\end{tabular}
}
\end{center}
\end{table}

\subsection{Wind speed}
\label{winddata}
\begin{figure}
\begin{center}
\includegraphics[width=\columnwidth]{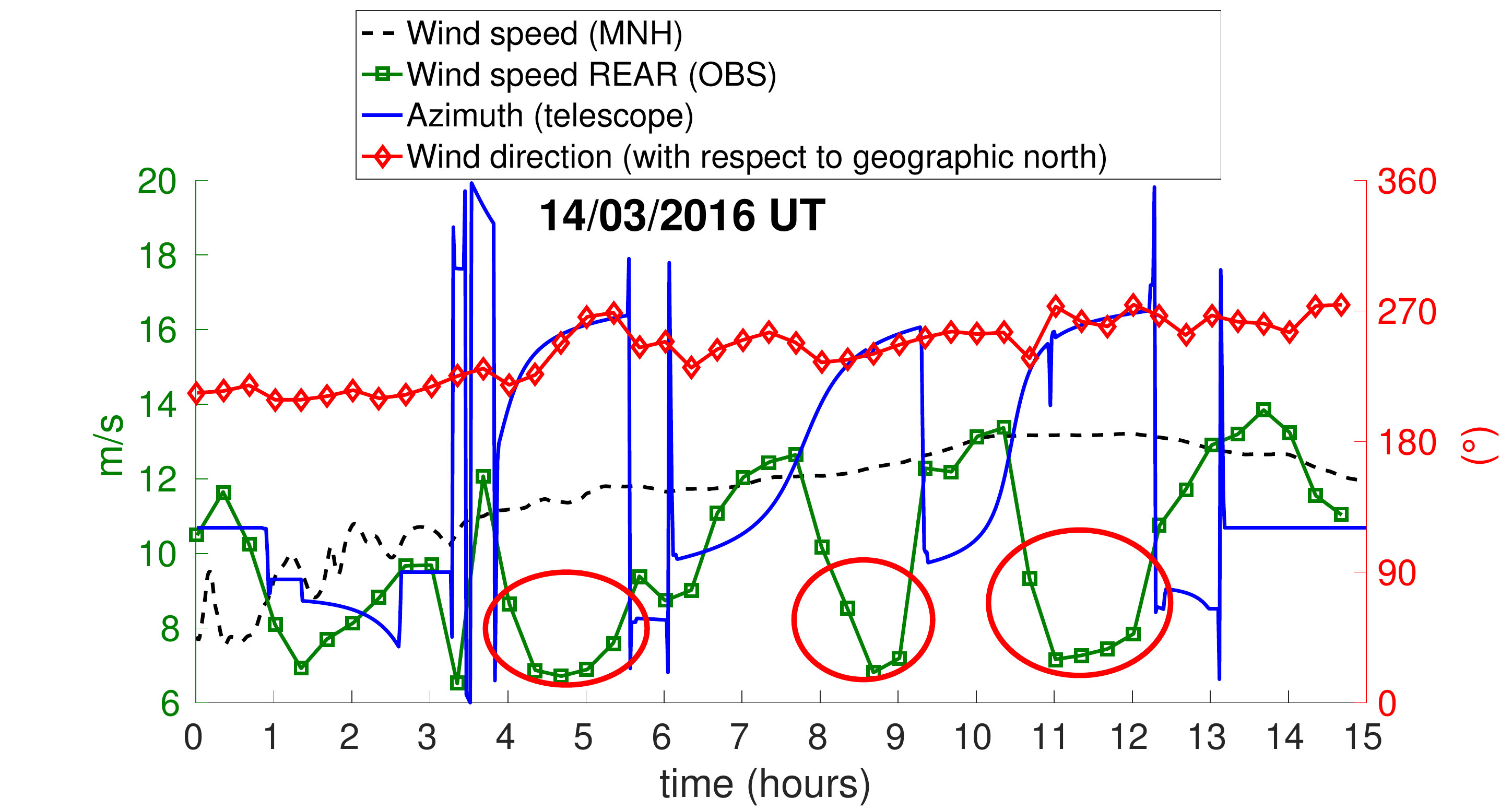}
\caption{Time evolution of different parameters along the test night of 14/03/2016 UT. The full blue line represents the telescope azimuth (right y-axis showing the angles), the full red line with diamond dots represents the incoming observed wind direction, the full green line with square dots represents the wind speed measured on the REAR anemometer, while the dashed black line represents the model output. The events in which the observed (REAR) wind speed value drops drastically are highlighted with red circles. These events coincide with the telescope azimuth facing the incoming wind direction.}
\label{angledir}
\end{center}
\end{figure}

\begin{figure*}
\begin{center}
\begin{tabular}{cc}
\includegraphics[height=5cm]{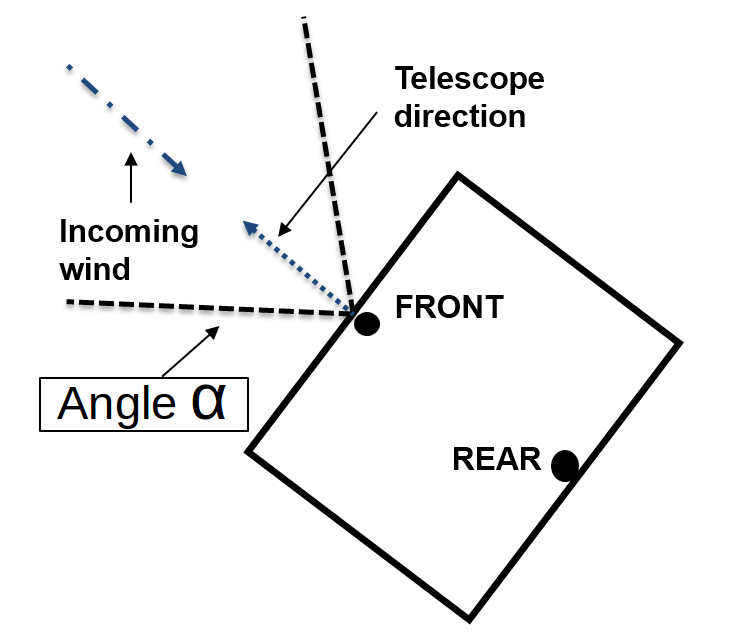} & \includegraphics[height=5cm]{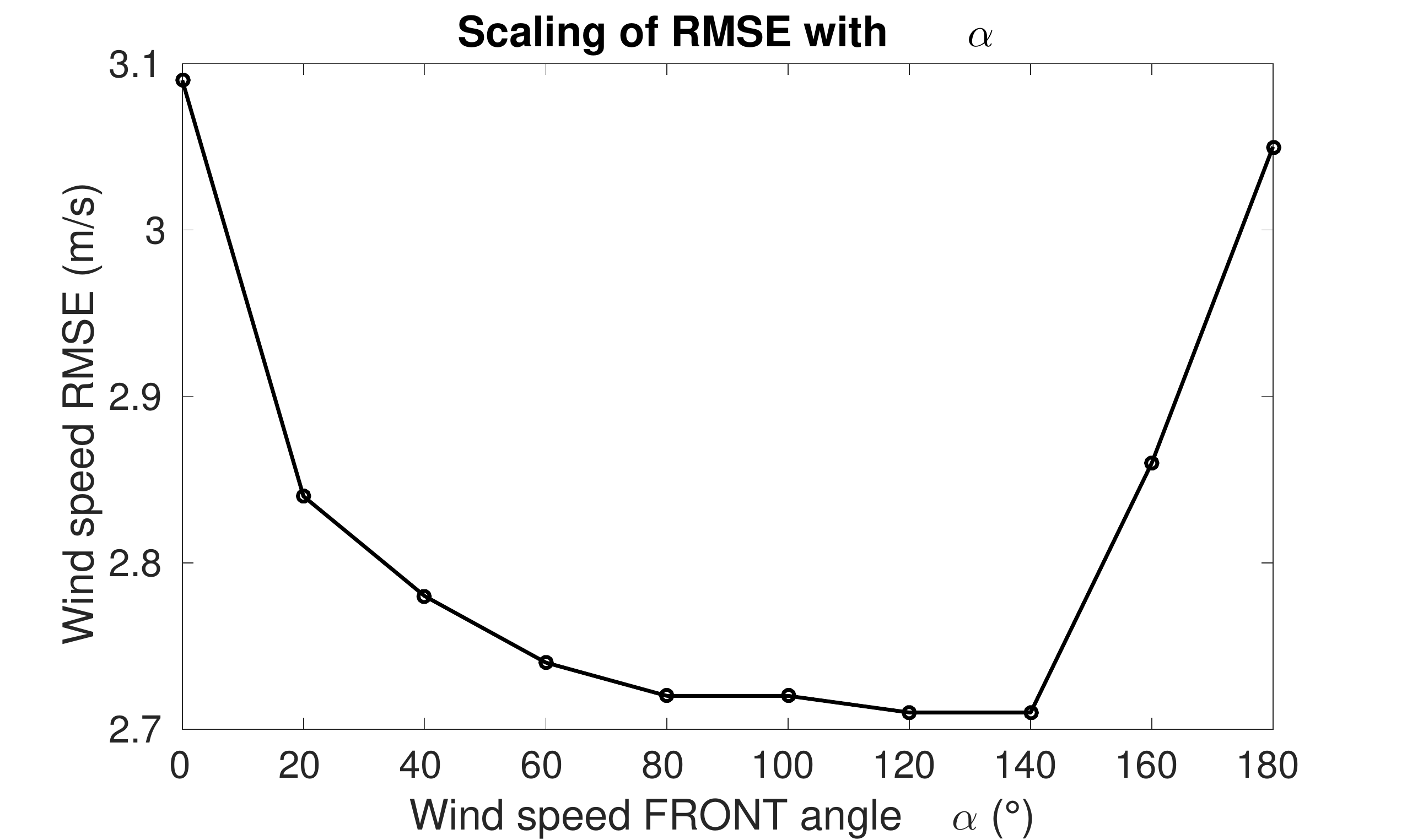}
\end{tabular}
\caption{\textbf{Left:} Wind speed selection scheme. Wind speed is taken from the FRONT anemometer if wind is coming within incident angle $\alpha$ with respect to the telescope orientation (azimuth), otherwise the measure is taken from the REAR anemometer. \textbf{Right:} RMSE between model outputs and wind speed measurements with respect to the total angle $\alpha$.}
\label{telescopedir}
\end{center}
\end{figure*}

While studying wind speed measurements, we discovered that FRONT and REAR anemometers tend to give significantly different measurements in some cases. Wind speed values tend to disagree when wind is coming from specific directions: if the wind is coming from the front of the telescope, the REAR anemometer tends to give a lower wind speed value with respect to the FRONT one, while the opposite is true if the wind is coming from the back of the telescope. We report in Fig. \ref{angledir} (left side) the observed wind speed behaviour for the test night of 14/03/2016 UT. It is evident that when the telescope orientation (azimuth) is coincident with the measured wind direction, so that the wind is coming from the front of the telescope, the wind speed measured on the REAR anemometer suddenly drops more than 5~ms$^{-1}$. Also, the wind speed predicted by the model tends to agree with the anemometer which is facing the incoming wind direction. We interpreted this evidence as the proof of a drag effect that slows down the wind passing over the dome.\\
Thanks to the above finding, we looked for a criterion to select wind speed measurements to be used in this validation study: the idea is that wind speed is taken from the FRONT anemometer if wind is coming within a specific incident angle $\alpha$ (measured by the telescope anemometers) with respect to the telescope orientation (azimuth), otherwise the measure is taken from the REAR anemometer (see Fig. \ref{telescopedir} - left side). To identify the width of this specific angle we selected different angle apertures from which to select FRONT anemometer measures and we looked for when the RMSE was minimum (see Fig. \ref{telescopedir} - right side). We observed that there is a saturation effect between an aperture of $\alpha=60^\circ$ ($\pm 30^\circ$ from the telescope azimuth) and $\alpha=140^\circ$ ($\pm 70^\circ$ from the telescope azimuth) where the RMSE, with respect to the model outputs, is minimum. We also observed that selecting any value of $\alpha$ within the range [60$^{\circ}$, 140$^{\circ}$] does not significantly change the values of the other statistical indicators used in this paper. For this reason, we decided to select FRONT wind speed measurements if the incident angle of the wind is within $\alpha=60^{\circ}$ with respect to the telescope azimuth, otherwise we select the REAR measurements. \\
 
\begin{figure*}
\begin{center}
\begin{tabular}{cc}
\includegraphics[width=5cm]{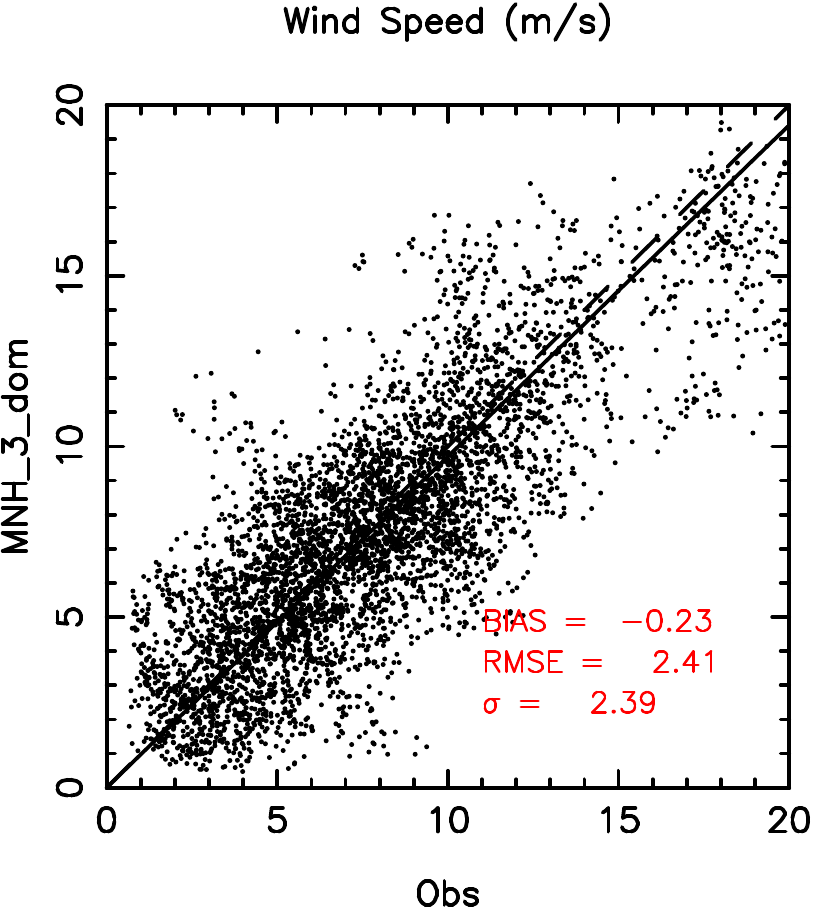} & \includegraphics[width=5cm]{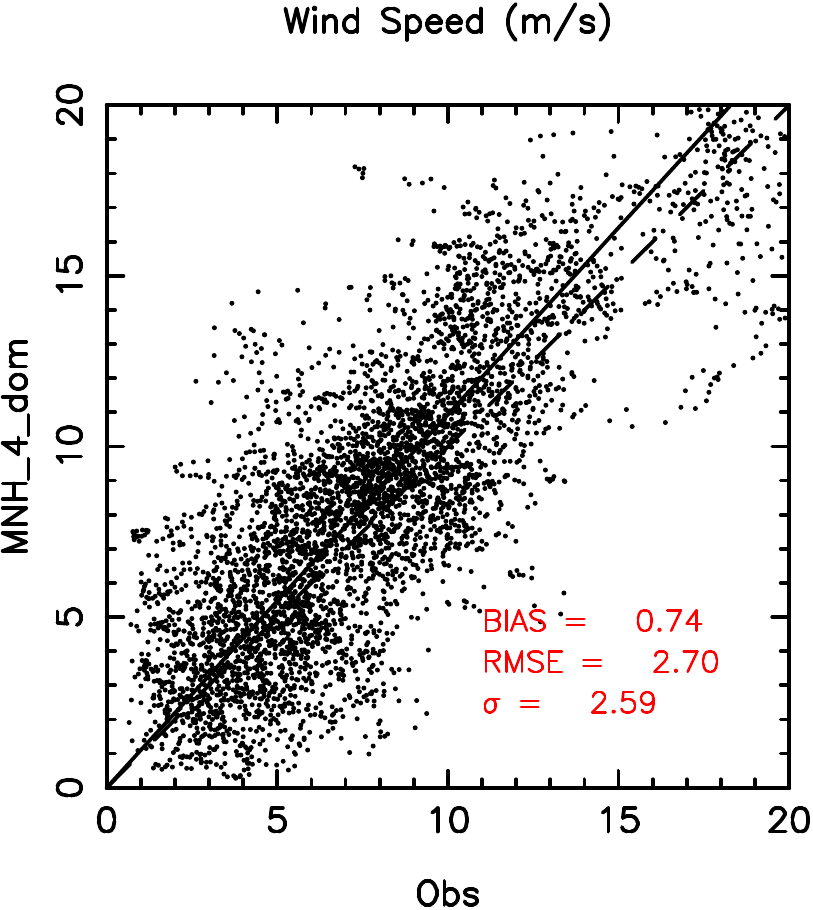}
\end{tabular}
\end{center}
\caption[example]{\label{fig:wsreg}Scatter plot for wind speed, comparing model outputs (MNH) and measurements (OBS). The full black line is the regression line passing by the origin, while the dashed line represent the reference diagonal line for unbiased results. Left: results obtained with the model output at 500~m horizontal resolution (domain 3). Right: results obtained with the model output at 100~m horizontal resolution (domain 4)}
\end{figure*}

While other parameters were all computed at 500~m horizontal resolution (see Table \ref{tab:resol}), in this section we report the results of the wind speed (measured with the criterion explained above) computed with two different horizontal resolutions on the sample reported in Table \ref{tab:samplesize}. In the left panel of Fig. \ref{fig:wsreg} we report the scatter plots computed with 500~m horizontal resolution, in the right panel the plot computed with 100~m horizontal resolution. We observe that the overall result is slightly better at 500~m resolution, with negligible bias = -0.2~ms$^{-1}$, RMSE = 2.4~ms$^{-1}$ and $\sigma$ = 2.4~ms$^{-1}$, while at 100~m resolution the bias=0.7~ms$^{-1}$ is slightly larger, yielding an RMSE = 2.7~ms$^{-1}$ and $\sigma$ = 2.6~ms$^{-1}$. However, as explained in section \ref{modelconf} and in \cite{lascaux2015}, it is evident that the results at 500~m resolution tend to underestimate strong wind speeds (larger than 10~ms$^{-1}$).\\
The above result is confirmed in the contingency tables \ref{tab:wsc} and \ref{tab:wsc2}, which reports the results obtained with 500~m and 100~m resolution respectively. In both cases we obtain a good global PC$\simeq$65\%. At 500~m resolution we have a higher performance at low-medium wind speeds, with POD$_1$=67.2\% and POD$_2$=57.9\%, while at 100~m resolution the performance is POD$_1$=61.0\% and POD$_2$=42.3\%. The result is reversed in the strong wind case, with a good POD$_3$=72.3\% at 500~m resolution that increases to an excellent POD$_3$=87.5\% at 100~m resolution. This is a logic result because it tells us that the higher resolution starts to play an important role when the wind speed is strong. We must also notice that in both cases the weaker performance reported in the POD$_2$ is influenced by the fact that the interval between the first and second tertile is 3.6~ms$^{-1}$ which is just slightly larger than to the RMSE values. \\

Since the most interesting application, from an astronomical point of view, is to correctly predict ground-layer strong winds that would affect telescope operations, we definitely need 100~m horizontal resolution to obtain the best performance. However, since low wind speeds are better modelled by a 500~m horizontal resolution model, in the operative setup we will end up using a mixed strategy of forecast, which will display results obtained with the 100~m resolution only in those nights characterized by strong wind speed. A test done on the sample we analyzed in this paper tells us that a good strategy should be to take as a threshold 8.5~ms$^{-1}$ (second tertile of the wind speed distribution in Fig. \ref{fig:tempcumdist1}). When the wind speed of the night is larger than this threshold we take the result of the model run at 100~m resolution, when the wind speed is lower we take results obtained by the model at 500~m resolution. In the operational procedure, we run, indeed, both models in sequence and we display the results from the 100~m or 500~m resolution wether the wind speed value from 100~m resolution  is larger or weaker than 8.5~ms$^{-1}$.
With the threshold at 8.5~ms$^{-1}$ we obtain a POD$_1$=65.7\%, POD$_2$=45.0\% and POD$_3$=84.8\% that provides us an excellent POD$_{3}$. Also for the wind speed we confirm the good model performances already observed above Cerro Paranal \citep{lascaux2015}. \\

If we consider a sample of bias, RMSE and $\sigma$ calculated for each night and we calculate the cumulative distribution (see Annex \ref{appendixcum}) we observe that the median values of the above parameters are lower than the the ones obtained on the whole sample. We obtain a median bias=0.2~ms$^{-1}$, a media RMSE=2.0~ms$^{-1}$, a median $\sigma$=1.6~ms$^{-1}$. From the distributions we observe that the nights in which the RMSE is larger than 4~ms$^{-1}$ are less than 5\% of the total, with rare events (less than 1\%) in which the RMSE is larger than 6~ms$^{-1}$.\\ 

Looking at the scattering pots of the sub-sample of summer and winter (Fig. \ref{fig:scattersummer} and \ref{fig:scatterwinter}, third panel) we do not observe any significant seasonal difference. Also the contingency tables (Table \ref{tab:wssummerc}, \ref{tab:wssummerc2}, \ref{tab:wswinterc2} and \ref{tab:wswinterc}) in summer and winter confirm an equivalent good behavior of the model with similar performances.\\

\begin{table}
\begin{center}
\caption{3$\times$3 contingency table for the wind speed during the night, at 58~m a.g.l. at LBT, for the sample of 139 nights.
We use the Meso-Nh ${\Delta}$X~=~500~m configuration.} 
\resizebox{\columnwidth}{!}{
\begin{tabular}{cc|ccc}
 \hline
 \multicolumn{2}{c}{Wind speed (ms$^{-1}$)} & \multicolumn{3}{c}{\bf OBSERVATIONS}\\
 \multicolumn{2}{c}{500~m res} & WS$<$4.9~ &4.9$<$WS$<$8.5  &  WS$>$8.5 \\
 \hline
 \multirow{7}{*}{\rotatebox{90}{\bf MODEL}} & & &\\
  & WS$<$4.9 &    847      &    388      &    24     \\
  & & & & \\
  &    4.9$<$WS$<$8.5 &    368      &    1036      &    515     \\
  & & & & \\
  & WS$>$   8.5 &    46      &    365      &    1407     \\
  & & & & \\
 \hline
 \\
 \multicolumn{5}{l}{Sample size = 4996; PC=65.9\%; EBD=1.4\%} \\
 \multicolumn{5}{l}{POD$_1$=67.2\%; POD$_2$=57.9\%; POD$_3$=72.3\%} \\
\end{tabular}
}
\label{tab:wsc}
\end{center}
\end{table}

\begin{table}
\begin{center}
\caption{3$\times$3 contingency table for the wind speed during the night, at 58~m a.g.l. at LBT, for the sample of 139 nights.
We use the Meso-Nh ${\Delta}$X~=~100~m configuration.} 
\resizebox{\columnwidth}{!}{
\begin{tabular}{cc|ccc}
 \hline
 \multicolumn{2}{c}{Wind speed (ms$^{-1}$)} & \multicolumn{3}{c}{\bf OBSERVATIONS}\\
 \multicolumn{2}{c}{100~m res} & WS$<$4.9 &4.9$<$WS$<$8.5  &  WS$>$8.5 \\
 \hline
 \multirow{7}{*}{\rotatebox{90}{\bf MODEL}} & & &\\
  & WS$<$4.9 &    769      &    382      &    9     \\
  & & & & \\
  &    4.9$<$WS$<$8.5 &    402      &    756      &    235     \\
  & & & & \\
  & WS$>$   8.5 &    90      &    651      &    1702     \\
  & & & & \\
 \hline
 \\
 \multicolumn{5}{l}{Sample size = 4996; PC=64.6\%; EBD=2.0\%} \\
 \multicolumn{5}{l}{POD$_1$=61.0\%; POD$_2$=42.3\%; POD$_3$=87.5\%} \\
\end{tabular}
}
\label{tab:wsc2}
\end{center}
\end{table}

\subsection{Wind direction}
\label{winddir}
In this analysis, we use the wind direction angle convention with respect to the geographical north, counting clockwise ($0^\circ$=North, $90^\circ$=East). We compared observed and simulated wind direction on different samples from which we filtered out all those measurements below a certain threshold, in order to compare model performances under different conditions. It is known indeed, that, when the wind speed is weak the dispersion of the wind direction is high and it is harder to reconstruct the right wind direction. If we think about the realistic scenarios, from an astronomical point of view we are more interested in knowing the accuracy  of the model in forecasting wind directions when the wind speed is high, because this impacts on telescope operations and instrument performances.\\
In this paper we decided to analyze the model performances with three different wind speed thresholds: 3~ms$^{-1}$, 5~ms$^{-1}$ and 10~ms$^{-1}$. The first threshold correspond to negligible wind speed values, the second one is more or less close to the median value of the wind speed (6.5 ms$^{-1}$) while the last threshold corresponds to moderately high wind speed values. The total sample size in nights for the wind direction is reported in Table \ref{tab:samplesize}, while in  Table \ref{tab:sizesun} is the total sample size in data points, once the wind speed filter is applied.\\

\begin{figure*}
\begin{center}
\begin{adjustbox}{max width=\textwidth}
\begin{tabular}{ccc}
\includegraphics[width=6cm]{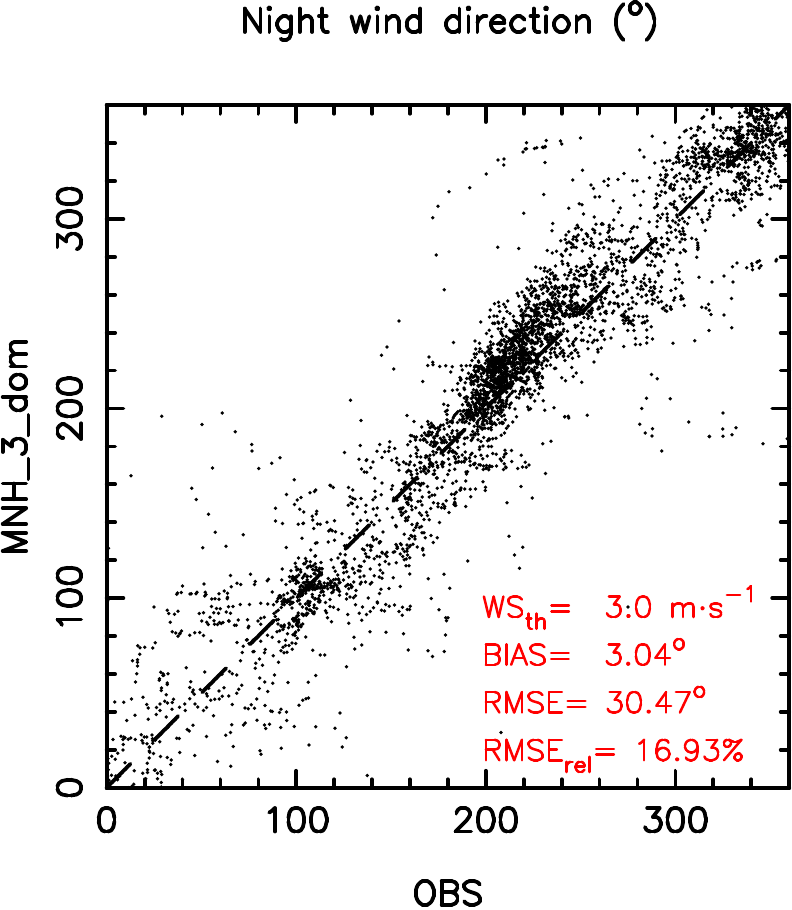} & \includegraphics[width=6cm]{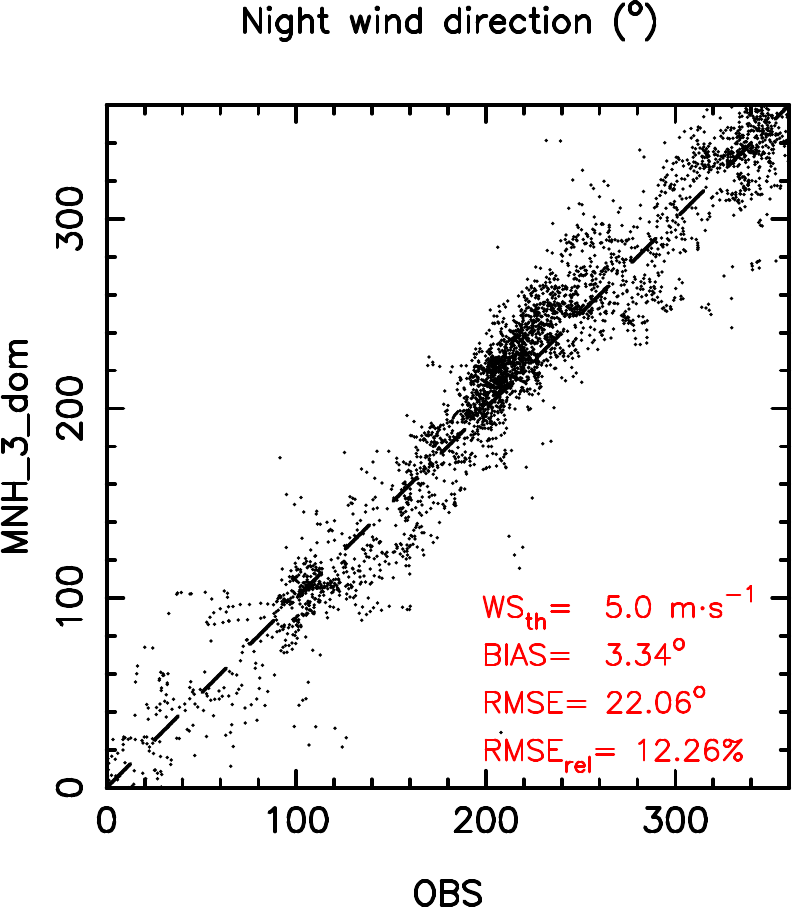} & \includegraphics[width=6cm]{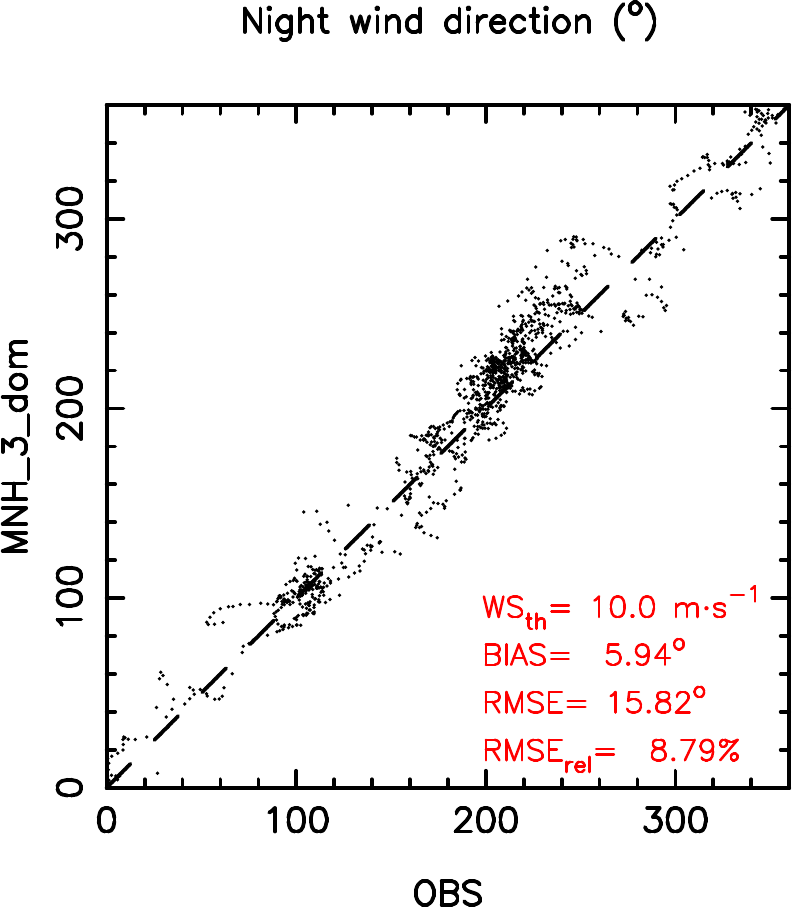}
\end{tabular}
\end{adjustbox}
\end{center}
\caption{Scatter plots for wind direction, comparing model outputs (MNH) and measurements (OBS). The dashed line represent the reference diagonal line for unbiased results. Wind directions are filtered by the corresponding value of the wind speed. Left: minimum wind speed of 3~ms$^{-1}$. Center: minimum wind speed of 5~ms$^{-1}$. Right minimum wind speed of 10~ms$^{-1}$.}
\label{fig:wdreg}
\end{figure*}

As introduced in \cite{lascaux2015}, we calculated another statistical quantity $RMSE_{rel} = \frac{RMSE}{180^\circ} \times 100$ i.e. the relative error of the RMSE calculated with respect the the worst possible prediction (reversed by $180^\circ$). \\

In Fig. \ref{fig:wdreg} we report the scatter plots obtained for the three different wind speed filters. In all cases results are extremely good with a maximum RMSE$_{rel}$ = 16.9\%. Besides, we observed that, as expected, the larger is the threshold on the wind speed, the better is the model performance. If we consider the threshold of 10 ms$^{-1}$ we find an excellent RMSE$_{rel}$ = 8.8\%.\\

Table \ref{tab:wdcon} report the contingency table computed for the first threshold (wind speed less than 3~ms$^{-1}$, the same used for Cerro Paranal) that permits us to discard just the very low wind speed values: the performance level is extremely good, with a global PC=75.4\% and PODs in the four wind quadrants ranging from 67.0\% to 81.5\%. In table \ref{tab:wdconsum} we show the computed contingency table statistical indicators with the three different wind speed filters. Raising the filter from 3~ms$^{-1}$ to 10~ms$^{-1}$ we see that EBD rapidly drops to zero, while PC raises up to 80.5\% for wind speeds greater than 10~ms$^{-1}$. In the case of strong winds we are able to obtain excellent PODs which ranges from 71.8\% to 97.5\%. If we consider that the wind speed comes mainly from South-West (see the wind rose Fig. \ref{fig:tempcumdist1}) we deduce that the model performances are within 71.8\% and 87.7\% if we consider the threshold at 10~ms$^{-1}$ and within 67\% and 80\% if we consider the threshold at  3~ms$^{-1}$.\\

The model seems to perform slightly worse for wind directions coming from South, however this is due to the fact the the wind rose distribution (Fig. \ref{fig:tempcumdist1}) shows that a significant portion of the winds come from an angle which is near the intersection of the West and South quadrants. If we rotate the division of quadrants of 45$^{\circ}$ and we consider South-East=$[90^\circ, 180^\circ]$ and South-West=$[180^\circ, 270^\circ]$ we obtain in these quadrants a POD of 76.6\% and 85.5\% respectively and a similar global PC=79.2\%, with a wind speed threshold of 3~ms$^{-1}$. The POD$_{SW}$ that corresponds to the sector with the highest frequency of the wind has therefore an excellent 85.5\%.

By observing the cumulative distributions of bias, RMSE and $\sigma$ computed over each single night, in Annex \ref{appendixcum}, computed with a wind speed threshold of 3~ms$^{-1}$ , we observe that the median RMSE=22$^\circ$ and $\sigma$=15$^\circ$ are much lower than in the ones obtained over the whole sample. 

By looking at the scattering pots of the sub-sample of summer and winter (Fig. \ref{fig:scattersummer} and \ref{fig:scatterwinter}, fourth panel, Annex \ref{appendix3}) we observe that the model performances are definitely better in winter time than in summer time. RMSE$_{rel}$=21\% in the summer season and an RMSE=13\% in the winter season. This is due to the fact that the winter season (see Fig. \ref{fig:tempcumdist3}) has much stronger winds than the summer season (see Fig. \ref{fig:tempcumdist2}) and the it is easier for the model to well reconstruct the wind direction when the wind speed is strong. This, as observed in table \ref{tab:wdconsum}, produces a statistics which is more favorable to model predictions.\\

\begin{table}
\begin{center}
\caption{4$\times$4 contingency table for the wind direction during the night, at 58m a.g.l. at LBT. We use the Meso-Nh ${\Delta}$X~=~500~m configuration. Wind speed (WS) threshold is 3~ms$^{-1}$.}
\begin{tabular}{cc|cccccc}
\multicolumn{2}{c}{Wind direction} & \multicolumn{4}{c}{\bf OBSERVATIONS}\\
\multicolumn{2}{c}{WS>3~ms$^{-1}$} & North & East & South & West \\
\hline
\multirow{9}{*}{\rotatebox{90}{\bf MODEL}} & & &\\
 & North & 737 & 53 & 20 & 145 \\
 &   & & & & \\
 & East & 63 & 517 & 125 & 5 \\
 & & & & & \\
 & South & 21 & 63 & 1177 & 74 \\
 &   & & & & \\
 & West & 85 & 1 & 434 & 903 \\
 &   & & & & \\
\hline
\\
\multicolumn{5}{l}{Sample size = 4423; PC=75.4\%; EBD=1.1\%} \\
\multicolumn{5}{l}{POD$_N$=81.3\%; POD$_E$=81.5\%} \\
\multicolumn{5}{l}{POD$_S$=67.0\%; POD$_W$=80.1\%} \\
\end{tabular}
\label{tab:wdcon}
\end{center}
\end{table}

\begin{table}
\begin{center}
\caption{Contingency table statistical indicators for the wind direction during the night computed with different wind speed filters, at 58m a.g.l. at LBT. We use the Meso-Nh ${\Delta}$X~=~500~m configuration.}
\begin{tabular}{c|ccc}
Operator & \multicolumn{3}{c}{Wind speed filters}\\
 & WS>3~ms$^{-1}$ & WS>5~ms$^{-1}$ & WS>10~ms$^{-1}$\\
\hline
PC & 75.4\% & 78.7\% & 80.5\%\\
EBD & 1.1\% & 0.1\% & 0.0\%\\
POD$_N$ & 81.3\% & 85.8\% & 85.5\%\\
POD$_E$ & 81.5\% & 89.2\% & 97.5\%\\
POD$_S$ & 67.0\% & 68.0\% & 71.8\%\\
POD$_W$ & 80.1\% & 85.1\% & 87.7\%\\
\hline
\end{tabular}
\label{tab:wdconsum}
\end{center}
\end{table}

\section{Conclusions}
\label{concl}
In this paper we presented the results of the validation study on the operational forecast system being developed for LBT, as part of the ALTA project, performed on a large 144 nights test sample uniformly distributed over the 2014-2015 solar years. The test was performed on the atmospheric parameters (temperature, relative humidity, wind speed and direction) near the ground level using, as a reference term, the telemetry data feed by the weather stations installed on the telescope dome. We used the Meso-Nh model with grid-nesting technique and an horizontal resolution of the innermost domain of 500~m and 100~m, the latter being essential in well reconstructing the wind speed in the case of strong winds. The results obtained in this validation study are extremely good and satisfactory and absolutely comparable to those obtained above Paranal and Armazones. Different statistical operators have been used (bias, RMSE and $\sigma$) and more sophisticated operators derived from the computation of the contingency tables i.e. the percent of correct detection (PC), the probability to detect the value of a parameter inside a specific range of values (POD) and the probability to have an extremely bad detection. The validation of the forecasting system refers to the operational configuration of the model used in the ALTA project. We summarize here the most relevant results:
\begin{enumerate}

\item The model has an excellent degree of reliability in reconstructing the temperature, with bias, RMSE and $\sigma$ below one degree Celsius.  We obtained an excellent PC=91.7$\%$ and all POD$_{i}$ are between 84$\%$ and 99$\%$. \\

\item The model shows good performances in reconstructing the relative humidity. We find a very satisfactory bias (bias=-2.36$\%$), RMSE$\simeq$14$\%$ and a $\sigma$ = 13.8$\%$. Besides the PC=71.8$\%$ is good too. The most critical POD$_{i}$ for the RH is POD$_{3}$ that is the probability to detect a RH larger than the second tertile. We obtained a POD$_{3}$ = 61\%. The model correctly discriminate the weak and the strong values of RH. We would like in the future to improve the dispersion in the every high values of RH but the present result is already more than satisfactory.\\

\item The model shows good performances in reconstructing the wind speed with an RMSE ranging from 2.4~ms$^{-1}$ to 2.7~ms$^{-1}$, depending on the horizontal resolution (500~m or 100~m). We observed that the resolution of 100~m is necessary only when the wind speed is large ($\gid$ 8.5 ms$^{-1}$) otherwise 500~m can provide even better results. We conceived therefore a hybrid treatment of the model outputs that consider the model outputs from the run having the highest resolution of 500~m when the average of the wind speed of the night is below 8.5 ms$^{-1}$ and a resolution of 100~m when it is larger than 8.5 ms$^{-1}$. With such a hybrid configuration we obtained a very satisfactory PC$\simeq$65$\%$, POD$_{1}$ = 65.7\%, POD$_{2}$ = 45$\%$ and POD$_{3}$ = 84.8$\%$. We highlight that POD$_{3}$ is the most important result for observational applications, being the strong wind case the critical one for telescope operations. We conclude that model performance is excellent in this respect.\\

\item Also the results obtained for the wind direction are very satisfactory, with an RMSE$_{rel}$=16.9$\%$ if we consider all data with wind speeds greater than 3~ms$^{-1}$. Besides we observed that RMSE$_{rel}$ can arrive to 8.8$\%$ if we consider all data with wind speed larger than 10~ms$^{-1}$. This means that the higher is the wind speed, the better is the reconstruction of the wind direction by the model. Values of PC and POD$_{i}$ improve also passing from a filtering of 3~ms$^{-1}$ to 10~ms$^{-1}$ with the best PC=80.5$\%$. POD$_{i}$ in all the four quadrants are excellent with values always larger than 72$\%$ and, when we filter out wind speed lower than 10~ms$^{-1}$, model performances achieve percents larger than 90$\%$ (see Table \ref{tab:wdconsum}). In particular, the POD$_{SW}$ related to the most interesting quadrant from which the wind comes more frequently (South-West) is an excellent 85.5$\%$. 

\end{enumerate}

Next step of our investigation for the ALTA Center project will be to use the Astro-Meso-Nh model in its most recent version \citet{masciadri2017} to forecast the optical turbulence at LBT. We will take advantage of measurements related to an extended site testing campaign \citet{masciadri2010} providing a vertical stratification of the optical turbulence on the whole 20 km above the ground in which a new technique for high-vertical resolution close to the ground \citet{egner2007} was employed. A study \citep{hagelin2011} done with a previous version of the Astro-Meso-Nh model, has already shown promising perspectives in that context. The Astro-Meso-Nh as well as the strategy evolved since there. The model requires therefore to re-clibrated and re-validated in this perspective.


\section*{Acknowledgements}

ALTA Center project is funded by the Large Binocular Telescope Corporation. Measurements of surface parameters are provided by the LBT telemetry and annexed instrumentation. The authors thanks Christian Veillet, Director of the Large Binocular Telescope, for his continued and valuable support given to this research activity. Authors also thanks the LBTO staff for their technical support and collaboration. Part of the numerical simulations have been run on the HPCF cluster of the European Center for Medium Range Weather Forecasts (ECMWF) using resources from the Project SPITFOT.








\newpage
\appendix

\section{Cumulative distributions}
\label{appendixcum}
\begin{figure*}
\centering
\includegraphics[width=0.9\textwidth]{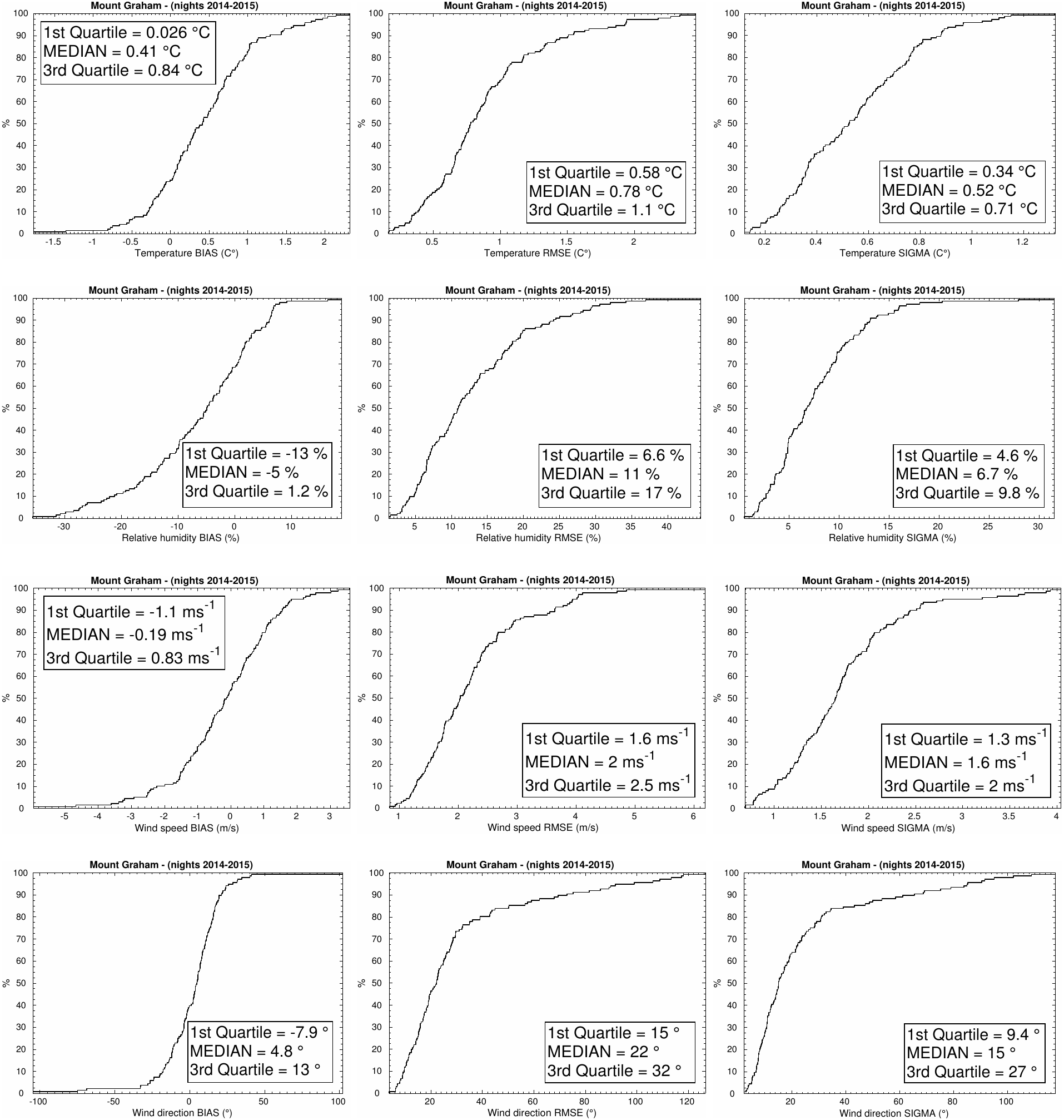}
\caption{Cumulative distributions of bias, RMSE and $\sigma$ for the atmospheric parameters. Results relative to the whole 2014-2015 years.}
\label{fig:tempcumdisterr}
\end{figure*}

\begin{figure*}
\centering
\includegraphics[width=0.9\textwidth]{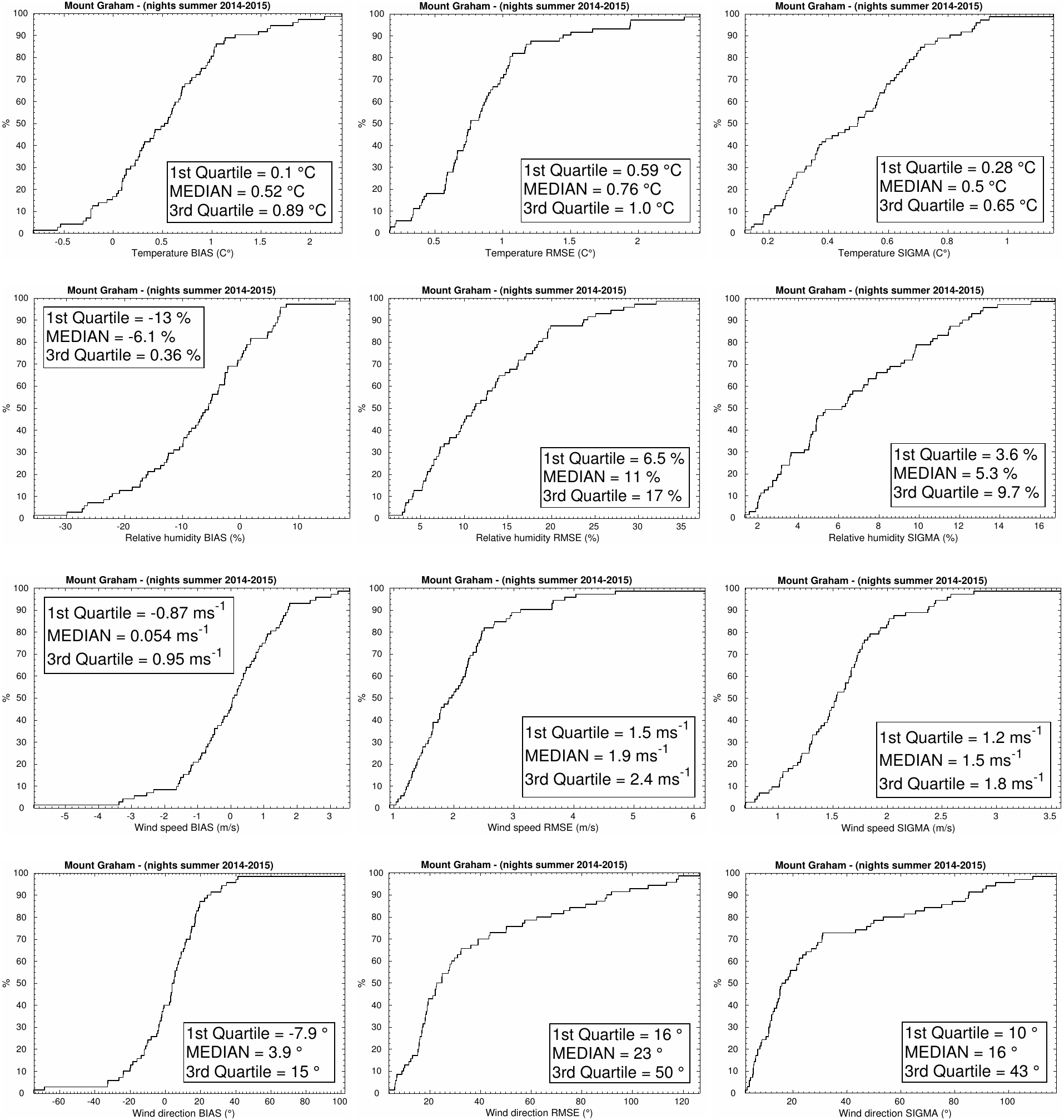}
\caption{Cumulative distributions of bias, RMSE and $\sigma$ for the atmospheric parameters. Results relative to the summers (April-September) of 2014-2015. Wind direction statistical operators are computed with a wind speed threshold of 3~ms$^{-1}$ (see section \ref{winddir}).}
\label{fig:tempcumdisterrsummer}
\end{figure*}

\begin{figure*}
\centering
\includegraphics[width=0.9\textwidth]{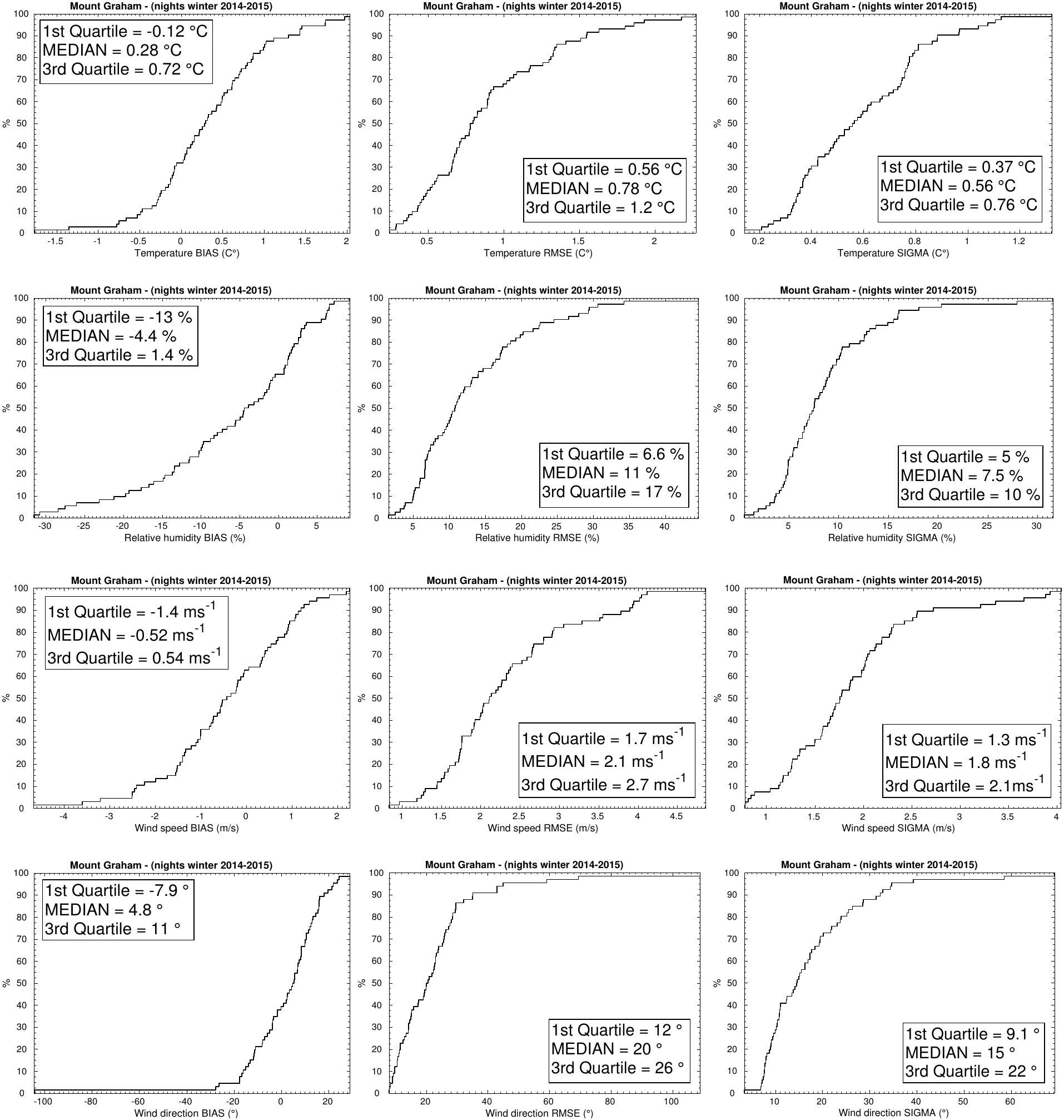}
\caption{Cumulative distributions of bias, RMSE and $\sigma$ for the atmospheric parameters. Results relative to the winters (October-March) of 2014-2015. Wind direction statistical operators are computed with a wind speed threshold of 3~ms$^{-1}$ (see section \ref{winddir}).}
\label{fig:tempcumdisterrwinter}
\end{figure*}

We report in this section the cumulative distributions of bias, RMSE and $\sigma$ for all atmospheric parameters at Mount Graham. Sample sizes for each variable are reported in Table \ref{tab:samplesize}. Wind direction statistical operators are computed with a wind speed threshold of 3~ms$^{-1}$ (see section \ref{winddir}). Values are computed over the whole two years 2014-2015 (Fig. \ref{fig:tempcumdisterr}), over the summer seasons of 2014-2015 (Fig. \ref{fig:tempcumdisterrsummer}) and over the winter seasons of the same years (Fig. \ref{fig:tempcumdisterrwinter}). 

\clearpage
\section{Seasonal validation}
\label{appendix3}
In this section we report the same validation study discussed in section \ref{validation}, performed on the subsample of nights from April to September (summer) and in the subsample from October to March (Winter). This is done in order to give the partial validation against the seasonal variation. We refer to the discussion about each parameter in section \ref{validation} for the specific methods used to validate the model. Here we report only the scatter plots and contingency tables obtained for each parameter in the specific seasonal subsample. Reported seasonal variations show a relevant difference for what regards wind direction, where the model performs better during winter time. This is eventually due to the fact that during winter wind speed is significantly higher (see figures \ref{fig:tempcumdist2} and \ref{fig:tempcumdist3}), which in turns reflect on the capability of the model in reconstructing the wind direction, as already discussed in section \ref{winddir}.

\begin{figure*}
\centering
\includegraphics[width=\textwidth]{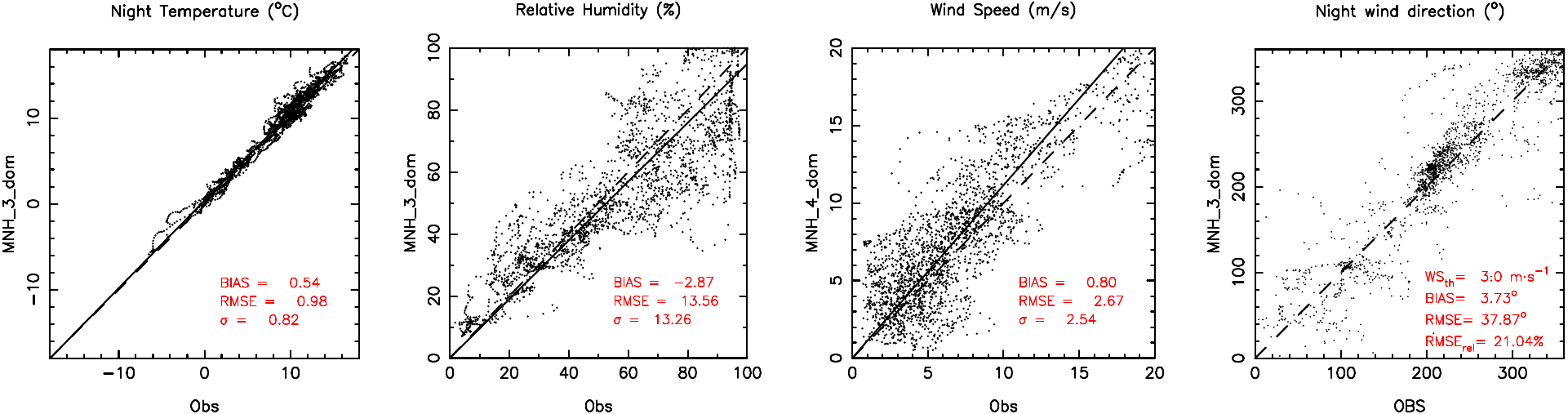}
\caption{Scatter plots computed for each parameter in the summer subsample (April-September). Wind speed is computed at 100~m horizontal resolution.}
\label{fig:scattersummer}
\end{figure*}

\begin{figure*}
\centering
\includegraphics[width=\textwidth]{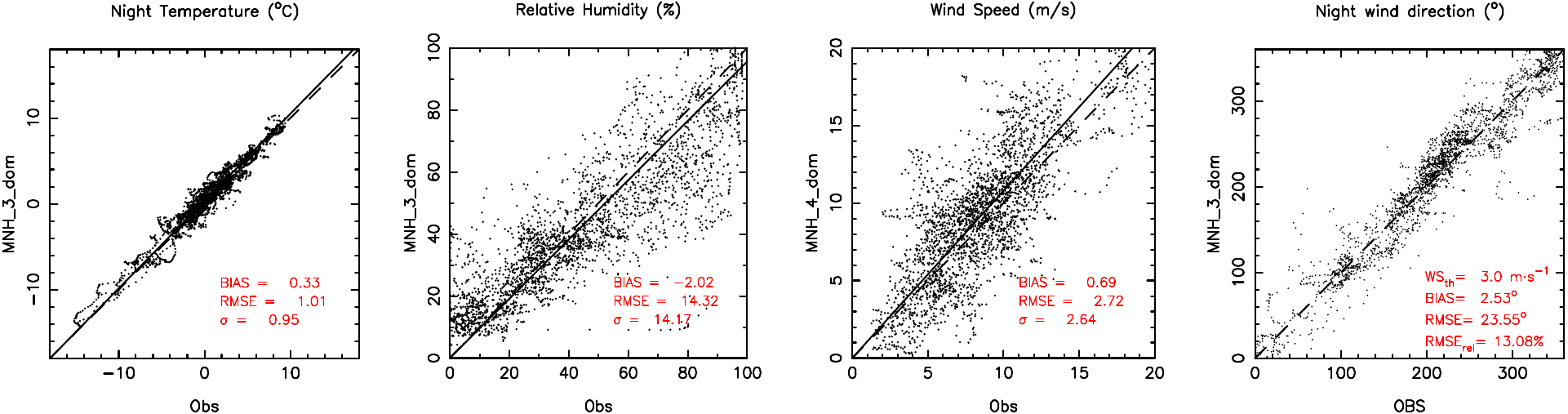}
\caption{Scatter plots computed for each parameter in the winter subsample (October-March). Wind speed is computed at 100~m horizontal resolution.}
\label{fig:scatterwinter}
\end{figure*}

\begin{table}
\begin{center}
\caption{}
\resizebox{\columnwidth}{!}{
\begin{tabular}{cc|cccccc}
 \hline
 \multicolumn{2}{c}{Temperature ($^{\circ}C$)} & \multicolumn{3}{c}{\bf OBSERVATIONS}\\
 \multicolumn{2}{c}{SUMMER} & ~~T$<$6.9 &    ~~6.9$<$T$<$10.7~~   &  ~~T$>$10.7 \\
 \hline
 \multirow{7}{*}{\rotatebox{90}{\bf MODEL}} & & &\\
  & T$<$6.9 &    771     &    10      &    0     \\
  & & & & \\
  &    6.9$<$T$<$10.7 &    10    &    538      &    63     \\
  & & & & \\
  & T$>$10.7 &    0 &    219    &    704     \\
  & & & & \\
 \hline
 \\
 \multicolumn{5}{l}{Sample size = 2315; PC=87.0\%; EBD=0.0\%} \\
 \multicolumn{5}{l}{POD$_1$=98.7\%; POD$_2$=70.1\%; POD$_3$=91.8\%} \\
\end{tabular}
}
\label{tab:tempsummerc}
\end{center}
\end{table}

\begin{table}
\begin{center}
\caption{}
\resizebox{\columnwidth}{!}{
\begin{tabular}{cc|cccccc}
 \hline
 \multicolumn{2}{c}{Relative humidity (\%)} & \multicolumn{3}{c}{\bf OBSERVATIONS}\\
 \multicolumn{2}{c}{SUMMER} & RH$<$37.7 &    37.7$<$RH$<$67.4 &  RH$>$67.4 \\
 \hline
 \multirow{7}{*}{\rotatebox{90}{\bf MODEL}} & & &\\
  & RH$<$37.7 &    562    &    78   &    4    \\
  & & & & \\
  &    37.7$<$RH$<$67.4 &    94   &    546    &    321    \\
  & & & & \\
  & RH$>$67.4 &    0      &    126     &    548    \\
  & & & & \\
 \hline
 \\
 \multicolumn{5}{l}{Sample size = 2279; PC=72.7\%; EBD=0.2\%} \\
  \multicolumn{5}{l}{POD$_1$=85.7\%; POD$_2$=72.8\%; POD$_3$=62.8\%} \\
\end{tabular}
}
\label{tab:rhsummerc}
\end{center}
\end{table}

\begin{table}
\begin{center}
\caption{}
\resizebox{\columnwidth}{!}{
\begin{tabular}{cc|cccccc}
 \hline
 \multicolumn{2}{c}{Wind speed (ms$^{-1}$)} & \multicolumn{3}{c}{\bf OBSERVATIONS}\\
 \multicolumn{2}{c}{SUMMER - 500~m res} & WS$<$4.2 &   4.2$<$WS$<$7.4  &  WS$>$7.4 \\
 \hline
 \multirow{7}{*}{\rotatebox{90}{\bf MODEL}} & & &\\
  & WS$<$   4.2 &    380   &    184    &    12  \\
  & & & & \\
  &    4.2 $<$WS$<$7.4 &    282     &    386      &    182     \\
  & & & & \\
  & WS$>$7.4 &    22      &    185      &    682     \\
  & & & & \\
 \hline
 \\
 \multicolumn{5}{l}{Sample size = 2315; PC=62.5\%; EBD=1.5\%} \\
  \multicolumn{5}{l}{POD$_1$=55.6\%; POD$_2$=51.1\%; POD$_3$=77.9\%} \\
\end{tabular}
}
\label{tab:wssummerc}
\end{center}
\end{table}

\begin{table}
\begin{center}
\caption{}
\resizebox{\columnwidth}{!}{
\begin{tabular}{cc|cccccc}
 \hline
 \multicolumn{2}{c}{Wind speed (ms$^{-1}$)} & \multicolumn{3}{c}{\bf OBSERVATIONS}\\
 \multicolumn{2}{c}{SUMMER - 100~m res} & WS$<$4.2 &   4.2$<$WS$<$7.4  &  WS$>$7.4 \\
 \hline
 \multirow{7}{*}{\rotatebox{90}{\bf MODEL}} & & &\\
  & WS$<$   4.2 &    327   &    191    &    22  \\
  & & & & \\
  &    4.2 $<$WS$<$7.4 &    299     &    324      &    57     \\
  & & & & \\
  & WS$>$7.4 &    58      &    240      &    797     \\
  & & & & \\
 \hline
 \\
 \multicolumn{5}{l}{Sample size = 2315; PC=62.5\%; EBD=3.5\%} \\
  \multicolumn{5}{l}{POD$_1$=47.8\%; POD$_2$=42.9\%; POD$_3$=91.0\%} \\
\end{tabular}
}
\label{tab:wssummerc2}
\end{center}
\end{table}

\begin{table}
\begin{center}
\caption{}
\resizebox{\columnwidth}{!}{
\begin{tabular}{cc|cccccc}
\multicolumn{2}{c}{Wind direction} & \multicolumn{4}{c}{\bf OBSERVATIONS}\\
\multicolumn{2}{c}{WS>3~ms$^{-1}$ - SUMMER} & North & East & South & West \\
\hline
\multirow{9}{*}{\rotatebox{90}{\bf MODEL}} & & &\\
 & North & 328  &   37    &   20    &   95 \\
 &   & & & & \\
 & East & 44 &  201  &   46   &   5 \\
 & & & & & \\
 & South & 16 &  24 &  516 &   31 \\
 &   & & & & \\
 & West & 33 &   1 & 181 &   301 \\
 &   & & & & \\
\hline
\\
\multicolumn{5}{l}{Sample size = 1879; PC=71.6\%; EBD=2.2\%} \\
\multicolumn{5}{l}{POD$_N$=77.9\%; POD$_E$=76.4\%} \\
\multicolumn{5}{l}{POD$_S$=67.6\%; POD$_W$=69.7\%} \\
\end{tabular}
}
\label{tab:wdsummerc}
\end{center}
\end{table}

\begin{table}
\begin{center}
\caption{}
\resizebox{\columnwidth}{!}{
\begin{tabular}{cc|cccccc}
 \hline
 \multicolumn{2}{c}{Temperature ($^{\circ}C$)} & \multicolumn{3}{c}{\bf OBSERVATIONS}\\
 \multicolumn{2}{c}{WINTER} & ~~T$<$-1.2 & ~~-1.2$<$T$<$2.4~~   &  ~~T$>$2.4 \\
 \hline
 \multirow{7}{*}{\rotatebox{90}{\bf MODEL}} & & &\\
  & T$<$-1.2 &    629 &    85      &    0     \\
  & & & & \\
  &   -1.2 $<$T$<$2.4 &    128      &    1044      &    88   \\
  & & & & \\
  & T$>$   2.4 &    0     &    135      &    777     \\
  & & & & \\
 \hline
 \\
 \multicolumn{5}{l}{Sample size = 2886; PC=84.9\%; EBD=0\%} \\
  \multicolumn{5}{l}{POD$_1$=83.1\%; POD$_2$=82.6\%; POD$_3$=89.9\%} \\
\end{tabular}
}
\label{tab:tempwinterc}
\end{center}
\end{table}

\begin{table}
\begin{center}
\caption{}
\resizebox{\columnwidth}{!}{
\begin{tabular}{cc|cccccc}
 \hline
 \multicolumn{2}{c}{Relative humidity(\%)} & \multicolumn{3}{c}{\bf OBSERVATIONS}\\
 \multicolumn{2}{c}{WINTER} & RH$<$24 &    24$<$RH$<$67.2  &  RH$>$67.2 \\
 \hline
 \multirow{7}{*}{\rotatebox{90}{\bf MODEL}} & & &\\
  & RH$<$24 &    624    &    94      &    13     \\
  & & & & \\
  &    24$<$RH$<$67.2 &    165    &    1103     &    324    \\
  & & & & \\
  & RH$>$67.2 &    0      &    70      &    493     \\
  & & & & \\
 \hline
 \\
 \multicolumn{5}{l}{Sample size = 2886; PC=76.9\%; EBD=0.5\%} \\
  \multicolumn{5}{l}{POD$_1$=79.1\%; POD$_2$=87.1\%; POD$_3$=59.4\%} \\
\end{tabular}
}
\label{tab:rhwinterc}
\end{center}
\end{table}

\begin{table}
\begin{center}
\caption{}
\resizebox{\columnwidth}{!}{
\begin{tabular}{cc|cccccc}
 \hline
 \multicolumn{2}{c}{Wind speed (ms$^{-1}$)} & \multicolumn{3}{c}{\bf OBSERVATIONS}\\
 \multicolumn{2}{c}{WINTER - 500~m res} & WS$<$5.6 &  5.6$<$WS$<$9.4  &  WS$>$9.4 \\
 \hline
 \multirow{7}{*}{\rotatebox{90}{\bf MODEL}} & & &\\
  & WS$<$5.6 &    378    &    253      &    26    \\
  & & & & \\
  &    5.6$<$WS$<$9.4 &    170      &    622      &    340    \\
  & & & & \\
  & WS$>$9.4 &    25    &    179     &    688   \\
  & & & & \\
 \hline
 \\
 \multicolumn{5}{l}{Sample size = 2681; PC=63.0\%; EBD=1.9\%} \\
  \multicolumn{5}{l}{POD$_1$=66.0\%; POD$_2$=59.0\%; POD$_3$= 65.3\%} \\
\end{tabular}
}
\label{tab:wswinterc2}
\end{center}
\end{table}

\begin{table}
\begin{center}
\caption{}
\resizebox{\columnwidth}{!}{
\begin{tabular}{cc|cccccc}
 \hline
 \multicolumn{2}{c}{Wind speed (ms$^{-1}$)} & \multicolumn{3}{c}{\bf OBSERVATIONS}\\
 \multicolumn{2}{c}{WINTER - 100~m res} & WS$<$5.6 &  5.6$<$WS$<$9.4  &  WS$>$9.4 \\
 \hline
 \multirow{7}{*}{\rotatebox{90}{\bf MODEL}} & & &\\
  & WS$<$5.6 &    338    &    201      &    11    \\
  & & & & \\
  &    5.6$<$WS$<$9.4 &    171      &    508      &    183    \\
  & & & & \\
  & WS$>$9.4 &    64    &    345     &    860   \\
  & & & & \\
 \hline
 \\
 \multicolumn{5}{l}{Sample size = 2681; PC=63.6\%; EBD=2.8\%} \\
  \multicolumn{5}{l}{POD$_1$=59.0\%; POD$_2$=48.2\%; POD$_3$= 81.6\%} \\
\end{tabular}
}
\label{tab:wswinterc}
\end{center}
\end{table}

\begin{table}
\begin{center}
\caption{}
\resizebox{\columnwidth}{!}{
\begin{tabular}{cc|cccccc}
\multicolumn{2}{c}{Wind direction} & \multicolumn{4}{c}{\bf OBSERVATIONS}\\
\multicolumn{2}{c}{WS>3~ms$^{-1}$ - WINTER} & North & East & South & West \\
\hline
\multirow{9}{*}{\rotatebox{90}{\bf MODEL}} & & &\\
 & North & 409 &   16 &  0 &  50 \\
 &   & & & & \\
 & East & 19 &     316 &    79 &     0 \\
 & & & & & \\
 & South & 5 &    39 &      661 &     43 \\
 &   & & & & \\
 & West & 52 &      0 &      253 &     602 \\
 &   & & & & \\
\hline
\\
\multicolumn{5}{l}{Sample size = 2544; PC=78.1\%; EBD=0.2\%} \\
\multicolumn{5}{l}{POD$_N$=84.3\%; POD$_E$=85.2\%} \\
\multicolumn{5}{l}{POD$_S$=66.6\%; POD$_W$=86.6\%} \\
\end{tabular}
}
\label{tab:wdwinterc}
\end{center}
\end{table}


\bsp	
\label{lastpage}
\end{document}